\documentclass[longauth]{aa}

\usepackage{color}
\usepackage{graphicx}
\usepackage{natbib}
\usepackage{enumitem}
\usepackage{txfonts}
\usepackage{multirow}
\usepackage{rotating}
\usepackage{longtable}
\usepackage{supertabular,booktabs}
\usepackage{amssymb}
\usepackage{verbatim}
\usepackage{caption}
\usepackage{subcaption}
\usepackage{marvosym}
\usepackage{textcomp}
\usepackage{hyperref}
\usepackage{euclid}
\usepackage{amsmath}
\usepackage{placeins}

\usepackage[utf8]{inputenc}

\titlerunning{C3R2 LBT}
\authorrunning{Saglia et al.}
\begin{document}

\title{\Euclid preparation: XX. The Complete Calibration of the
  Color--Redshift Relation survey: LBT observations and data release
\thanks{The LBT is an international collaboration among institutions in the United States, Italy, and Germany. The LBT Corporation partners are:  LBT Beteiligungsgesellschaft, Germany, representing the Max-Planck Society, the Astrophysical Institute Potsdam, and Heidelberg University; The University of Arizona on behalf of the Arizona university system; Istituto Nazionale di Astrofisica, Italy; The Ohio State University, and The Research Corporation, on behalf of The University of Notre Dame, University of Minnesota, and University of Virginia. }}


\author{Euclid Collaboration: R.~Saglia\orcid{0000-0003-0378-7032}$^{1,2}$\thanks{\email{saglia@mpe.mpg.de}}, S.~De~Nicola$^{1}$, M.~Fabricius$^{1}$, V.~Guglielmo$^{3}$, J.~Snigula$^{1}$, R.~Z\"oller$^{1,2}$, R.~Bender$^{1,2}$, J.~Heidt$^{4}$, D.~Masters$^{5}$, D.~Stern\orcid{0000-0003-2686-9241}$^{6}$
% ...
}

\institute{$^{1}$ Max Planck Institute for Extraterrestrial Physics, Giessenbachstr. 1, D-85748 Garching, Germany\\
$^{2}$ Universit\"ats-Sternwarte M\"unchen, Fakult\"at f\"ur Physik, Ludwig-Maximilians-Universit\"at M\"unchen, Scheinerstrasse 1, 81679 M\"unchen, Germany\\
$^{3}$ INAF-Osservatorio di Astrofisica e Scienza dello Spazio di Bologna, Via Piero Gobetti 93/3, I-40129 Bologna, Italy\\
$^{4}$ Landessternwarte, Zentrum f\"ur Astronomie der Universit\"at Heidelberg, K\"onigstuhl 12, 69117 Heidelberg, Germany\\
$^{5}$ Infrared Processing and Analysis Center, California Institute of Technology, Pasadena, CA 91125, USA\\
$^{6}$ Jet Propulsion Laboratory, California Institute of Technology, 4800 Oak Grove Drive, Pasadena, CA, 91109, USA\\
$^{7}$ Department of Astronomy, University of Geneva, ch. d\'Ecogia 16, CH-1290 Versoix, Switzerland\\
$^{8}$ Institute of Cosmology and Gravitation, University of Portsmouth, Portsmouth PO1 3FX, UK\\
$^{9}$ Dipartimento di Fisica e Astronomia, Universit\'a di Bologna, Via Gobetti 93/2, I-40129 Bologna, Italy\\
$^{10}$ INFN-Sezione di Bologna, Viale Berti Pichat 6/2, I-40127 Bologna, Italy\\
$^{11}$ INAF-Osservatorio Astrofisico di Torino, Via Osservatorio 20, I-10025 Pino Torinese (TO), Italy\\
$^{12}$ Dipartimento di Fisica, Universit\'a degli studi di Genova, and INFN-Sezione di Genova, via Dodecaneso 33, I-16146, Genova, Italy\\
$^{13}$ INFN-Sezione di Roma Tre, Via della Vasca Navale 84, I-00146, Roma, Italy\\
$^{14}$ INAF-Osservatorio Astronomico di Capodimonte, Via Moiariello 16, I-80131 Napoli, Italy\\
$^{15}$ Instituto de Astrof\'isica e Ci\^encias do Espa\c{c}o, Universidade do Porto, CAUP, Rua das Estrelas, PT4150-762 Porto, Portugal\\
$^{16}$ Dipartimento di Fisica, Universit\'a degli Studi di Torino, Via P. Giuria 1, I-10125 Torino, Italy\\
$^{17}$ INFN-Sezione di Torino, Via P. Giuria 1, I-10125 Torino, Italy\\
$^{18}$ INAF-IASF Milano, Via Alfonso Corti 12, I-20133 Milano, Italy\\
$^{19}$ Institut de F\'{i}sica d'Altes Energies (IFAE), The Barcelona Institute of Science and Technology, Campus UAB, 08193 Bellaterra (Barcelona), Spain\\
$^{20}$ Port d'Informaci\'{o} Cient\'{i}fica, Campus UAB, C. Albareda s/n, 08193 Bellaterra (Barcelona), Spain\\
$^{21}$ INAF-Osservatorio Astronomico di Roma, Via Frascati 33, I-00078 Monteporzio Catone, Italy\\
$^{22}$ INFN section of Naples, Via Cinthia 6, I-80126, Napoli, Italy\\
$^{23}$ Department of Physics "E. Pancini", University Federico II, Via Cinthia 6, I-80126, Napoli, Italy\\
$^{24}$ Institut national de physique nucl\'eaire et de physique des particules, 3 rue Michel-Ange, 75794 Paris C\'edex 16, France\\
$^{25}$ Institute for Astronomy, University of Edinburgh, Royal Observatory, Blackford Hill, Edinburgh EH9 3HJ, UK\\
$^{26}$ Jodrell Bank Centre for Astrophysics, Department of Physics and Astronomy, University of Manchester, Oxford Road, Manchester M13 9PL, UK\\
$^{27}$ ESAC/ESA, Camino Bajo del Castillo, s/n., Urb. Villafranca del Castillo, 28692 Villanueva de la Ca\~nada, Madrid, Spain\\
$^{28}$ European Space Agency/ESRIN, Largo Galileo Galilei 1, 00044 Frascati, Roma, Italy\\
$^{29}$ Univ Lyon, Univ Claude Bernard Lyon 1, CNRS/IN2P3, IP2I Lyon, UMR 5822, F-69622, Villeurbanne, France\\
$^{30}$ Institute of Physics, Laboratory of Astrophysics, Ecole Polytechnique F\'{e}d\'{e}rale de Lausanne (EPFL), Observatoire de Sauverny, 1290 Versoix, Switzerland\\
$^{31}$ Mullard Space Science Laboratory, University College London, Holmbury St Mary, Dorking, Surrey RH5 6NT, UK\\
$^{32}$ Departamento de F\'isica, Faculdade de Ci\^encias, Universidade de Lisboa, Edif\'icio C8, Campo Grande, PT1749-016 Lisboa, Portugal\\
$^{33}$ Instituto de Astrof\'isica e Ci\^encias do Espa\c{c}o, Faculdade de Ci\^encias, Universidade de Lisboa, Campo Grande, PT-1749-016 Lisboa, Portugal\\
$^{34}$ Universit\'e Paris-Saclay, CNRS, Institut d'astrophysique spatiale, 91405, Orsay, France\\
$^{35}$ Department of Physics, Oxford University, Keble Road, Oxford OX1 3RH, UK\\
$^{36}$ INFN-Padova, Via Marzolo 8, I-35131 Padova, Italy\\
$^{37}$ AIM, CEA, CNRS, Universit\'{e} Paris-Saclay, Universit\'{e} de Paris, F-91191 Gif-sur-Yvette, France\\
$^{38}$ INAF-Osservatorio Astronomico di Trieste, Via G. B. Tiepolo 11, I-34143 Trieste, Italy\\
$^{39}$ Aix-Marseille Univ, CNRS/IN2P3, CPPM, Marseille, France\\
$^{40}$ Istituto Nazionale di Astrofisica (INAF) - Osservatorio di Astrofisica e Scienza dello Spazio (OAS), Via Gobetti 93/3, I-40127 Bologna, Italy\\
$^{41}$ Istituto Nazionale di Fisica Nucleare, Sezione di Bologna, Via Irnerio 46, I-40126 Bologna, Italy\\
$^{42}$ INAF-Osservatorio Astronomico di Padova, Via dell'Osservatorio 5, I-35122 Padova, Italy\\
$^{43}$ Institute of Theoretical Astrophysics, University of Oslo, P.O. Box 1029 Blindern, N-0315 Oslo, Norway\\
$^{44}$ Leiden Observatory, Leiden University, Niels Bohrweg 2, 2333 CA Leiden, The Netherlands\\
$^{45}$ von Hoerner \& Sulger GmbH, Schlo{\ss}Platz 8, D-68723 Schwetzingen, Germany\\
$^{46}$ Technical University of Denmark, Elektrovej 327, 2800 Kgs. Lyngby, Denmark\\
$^{47}$ Max-Planck-Institut f\"ur Astronomie, K\"onigstuhl 17, D-69117 Heidelberg, Germany\\
$^{48}$ Universit\'e de Gen\`eve, D\'epartement de Physique Th\'eorique and Centre for Astroparticle Physics, 24 quai Ernest-Ansermet, CH-1211 Gen\`eve 4, Switzerland\\
$^{49}$ Department of Physics and Helsinki Institute of Physics, Gustaf H\"allstr\"omin katu 2, 00014 University of Helsinki, Finland\\
$^{50}$ European Space Agency/ESTEC, Keplerlaan 1, 2201 AZ Noordwijk, The Netherlands\\
$^{51}$ NOVA optical infrared instrumentation group at ASTRON, Oude Hoogeveensedijk 4, 7991PD, Dwingeloo, The Netherlands\\
$^{52}$ Argelander-Institut f\"ur Astronomie, Universit\"at Bonn, Auf dem H\"ugel 71, 53121 Bonn, Germany\\
$^{53}$ Dipartimento di Fisica e Astronomia "Augusto Righi" - Alma Mater Studiorum Universit\`{a} di Bologna, via Piero Gobetti 93/2, I-40129 Bologna, Italy\\
$^{54}$ Department of Physics, Institute for Computational Cosmology, Durham University, South Road, DH1 3LE, UK\\
$^{55}$ Institut d'Astrophysique de Paris, UMR 7095, CNRS, and Sorbonne Universit\'e, 98 bis boulevard Arago, 75014 Paris, France\\
$^{56}$ University of Applied Sciences and Arts of Northwestern Switzerland, School of Engineering, 5210 Windisch, Switzerland\\
$^{57}$ Department of Physics and Astronomy, University of Aarhus, Ny Munkegade 120, DK-8000 Aarhus C, Denmark\\
$^{58}$ Centre for Astrophysics, University of Waterloo, Waterloo, Ontario N2L 3G1, Canada\\
$^{59}$ Department of Physics and Astronomy, University of Waterloo, Waterloo, Ontario N2L 3G1, Canada\\
$^{60}$ Perimeter Institute for Theoretical Physics, Waterloo, Ontario N2L 2Y5, Canada\\
$^{61}$ Universit\'e Paris-Saclay, Universit\'e Paris Cit\'e, CEA, CNRS, Astrophysique, Instrumentation et Mod\'elisation Paris-Saclay, 91191 Gif-sur-Yvette, France\\
$^{62}$ Centre National d'Etudes Spatiales, Toulouse, France\\
$^{63}$ Institute of Space Science, Bucharest, Ro-077125, Romania\\
$^{64}$ Dipartimento di Fisica e Astronomia "G.Galilei", Universit\'a di Padova, Via Marzolo 8, I-35131 Padova, Italy\\
$^{65}$ Departamento de F\'isica, FCFM, Universidad de Chile, Blanco Encalada 2008, Santiago, Chile\\
$^{66}$ IFPU, Institute for Fundamental Physics of the Universe, via Beirut 2, 34151 Trieste, Italy\\
$^{67}$ Centro de Investigaciones Energ\'eticas, Medioambientales y Tecnol\'ogicas (CIEMAT), Avenida Complutense 40, 28040 Madrid, Spain\\
$^{68}$ Instituto de Astrof\'isica e Ci\^encias do Espa\c{c}o, Faculdade de Ci\^encias, Universidade de Lisboa, Tapada da Ajuda, PT-1349-018 Lisboa, Portugal\\
$^{69}$ Universidad Polit\'ecnica de Cartagena, Departamento de Electr\'onica y Tecnolog\'ia de Computadoras, 30202 Cartagena, Spain\\
$^{70}$ Kapteyn Astronomical Institute, University of Groningen, PO Box 800, 9700 AV Groningen, The Netherlands\\
$^{71}$ INAF-Osservatorio Astronomico di Brera, Via Brera 28, I-20122 Milano, Italy\\
$^{72}$ Dipartimento di Fisica "Aldo Pontremoli", Universit\'a degli Studi di Milano, Via Celoria 16, I-20133 Milano, Italy\\
$^{73}$ INFN-Sezione di Milano, Via Celoria 16, I-20133 Milano, Italy\\
$^{74}$ Dipartimento di Fisica e Astronomia "Augusto Righi" - Alma Mater Studiorum Universit\'a di Bologna, Viale Berti Pichat 6/2, I-40127 Bologna, Italy\\
$^{75}$ INFN, Sezione di Trieste, Via Valerio 2, I-34127 Trieste TS, Italy\\
$^{76}$ SISSA, International School for Advanced Studies, Via Bonomea 265, I-34136 Trieste TS, Italy\\
$^{77}$ Instituto de Astrof\'isica de Canarias (IAC); Departamento de Astrof\'isica, Universidad de La Laguna (ULL), E-38200, La Laguna, Tenerife, Spain\\
$^{78}$ Instituto de Astrof\'isica de Canarias, Calle V\'ia L\'actea s/n, E-38204, San Crist\'obal de La Laguna, Tenerife, Spain\\
$^{79}$ INFN-Bologna, Via Irnerio 46, I-40126 Bologna, Italy\\
$^{80}$ Dipartimento di Fisica e Scienze della Terra, Universit\'a degli Studi di Ferrara, Via Giuseppe Saragat 1, I-44122 Ferrara, Italy\\
$^{81}$ INAF, Istituto di Radioastronomia, Via Piero Gobetti 101, I-40129 Bologna, Italy\\
$^{82}$ Institut d'Astrophysique de Paris, 98bis Boulevard Arago, F-75014, Paris, France\\
$^{83}$ Universit\'e C\^{o}te d'Azur, Observatoire de la C\^{o}te d'Azur, CNRS, Laboratoire Lagrange, Bd de l'Observatoire, CS 34229, 06304 Nice cedex 4, France\\
$^{84}$ Institute for Theoretical Particle Physics and Cosmology (TTK), RWTH Aachen University, D-52056 Aachen, Germany\\
$^{85}$ Department of Physics \& Astronomy, University of California Irvine, Irvine CA 92697, USA\\
$^{86}$ University of Lyon, UCB Lyon 1, CNRS/IN2P3, IUF, IP2I Lyon, France\\
$^{87}$ INFN-Sezione di Genova, Via Dodecaneso 33, I-16146, Genova, Italy\\
$^{88}$ INAF-Istituto di Astrofisica e Planetologia Spaziali, via del Fosso del Cavaliere, 100, I-00100 Roma, Italy\\
$^{89}$ School of Physics, HH Wills Physics Laboratory, University of Bristol, Tyndall Avenue, Bristol, BS8 1TL, UK\\
$^{90}$  Universit\'e Paris Cit\'e, CNRS, Astroparticule et Cosmologie, F-75013 Paris, France\\
$^{91}$ Departamento de F\'isica Te\'orica, Facultad de Ciencias, Universidad Aut\'onoma de Madrid, 28049 Cantoblanco, Madrid, Spain\\
$^{92}$ Instituto de F\'isica Te\'orica UAM-CSIC, Campus de Cantoblanco, E-28049 Madrid, Spain\\
$^{93}$ Department of Physics, P.O. Box 64, 00014 University of Helsinki, Finland\\
$^{94}$ Ruhr University Bochum, Faculty of Physics and Astronomy, Astronomical Institute (AIRUB), German Centre for Cosmological Lensing (GCCL), 44780 Bochum, Germany\\
$^{95}$ Department of Physics, Lancaster University, Lancaster, LA1 4YB, UK\\
$^{96}$ Aix-Marseille Univ, CNRS, CNES, LAM, Marseille, France\\
$^{97}$ Code 665, NASA Goddard Space Flight Center, Greenbelt, MD 20771 and SSAI, Lanham, MD 20770, USA\\
$^{98}$ Astrophysics Group, Blackett Laboratory, Imperial College London, London SW7 2AZ, UK\\
$^{99}$ Department of Physics and Astronomy, University College London, Gower Street, London WC1E 6BT, UK\\
$^{100}$ Univ. Grenoble Alpes, CNRS, Grenoble INP, LPSC-IN2P3, 53, Avenue des Martyrs, 38000, Grenoble, France\\
$^{101}$ Centre de Calcul de l'IN2P3, 21 avenue Pierre de Coubertin F-69627 Villeurbanne Cedex, France\\
$^{102}$ Dipartimento di Fisica, Sapienza Universit\`a di Roma, Piazzale Aldo Moro 2, I-00185 Roma, Italy\\
$^{103}$ Department of Mathematics and Physics E. De Giorgi, University of Salento, Via per Arnesano, CP-I93, I-73100, Lecce, Italy\\
$^{104}$ INAF-Sezione di Lecce, c/o Dipartimento Matematica e Fisica, Via per Arnesano, I-73100, Lecce, Italy\\
$^{105}$ INFN, Sezione di Lecce, Via per Arnesano, CP-193, I-73100, Lecce, Italy\\
$^{106}$ Institute for Computational Science, University of Zurich, Winterthurerstrasse 190, 8057 Zurich, Switzerland\\
$^{107}$ Higgs Centre for Theoretical Physics, School of Physics and Astronomy, The University of Edinburgh, Edinburgh EH9 3FD, UK\\
$^{108}$ Universit\'e St Joseph; Faculty of Sciences, Beirut, Lebanon\\
$^{109}$ Institut de Recherche en Astrophysique et Plan\'etologie (IRAP), Universit\'e de Toulouse, CNRS, UPS, CNES, 14 Av. Edouard Belin, F-31400 Toulouse, France\\
$^{110}$ Department of Astrophysical Sciences, Peyton Hall, Princeton University, Princeton, NJ 08544, USA\\
$^{111}$ Department of Physics, P.O.Box 35 (YFL), 40014 University of Jyv\"askyl\"a, Finland\\
$^{112}$ Helsinki Institute of Physics, Gustaf H{\"a}llstr{\"o}min katu 2, University of Helsinki, Helsinki, Finland\\
  $^{113}$ Department of Mathematics and Physics, Roma Tre University, Via della Vasca Navale 84, I-00146 Rome, Italy}

\abstract{The Complete Calibration of the Color--Redshift Relation
  survey (C3R2) is a spectroscopic program  designed to empirically
  calibrate the galaxy color--redshift relation to the {\it Euclid}
  depth ($I_{\scriptscriptstyle{\rm E}}=24.5$), a key ingredient for
  the success of Stage IV dark energy projects based on weak lensing
  cosmology. A spectroscopic calibration sample that is as representative as
  possible of the galaxies in the {\it Euclid} weak lensing sample is
  being collected, selecting galaxies from a self-organizing map (SOM)
  representation of the galaxy color space. Here, we present the
  results of a near-infrared $H$- and $K$-band spectroscopic campaign
  carried out using the LUCI instruments at the LBT. For a total of
  251 galaxies, we present new highly reliable redshifts in the
  $1.3\leq z\leq 1.7 $ and $2\leq z\leq 2.7 $ ranges. The
  newly-determined redshifts populate 49 SOM cells that previously
  contained no spectroscopic measurements and almost twice the
  occupation numbers of an additional 153 SOM cells. A final optical
  ground-based observational effort is needed to calibrate the missing
  cells, in particular in the redshift range $1.7\leq z\leq 2.7$, which t
  lack spectroscopic calibration. In the end, {\it Euclid} itself will 
  deliver telluric-free near-IR spectra that can complete the
  calibration.}


\date{Received xxx; accepted yyy}

\maketitle

\section{Introduction}

The {\it Euclid} satellite \citep{Laureijs2011} is scheduled for
launch in 2023; it  will observe galaxies to $z>2$ over $15\,000\,\rm
deg^2$ using two instruments: VIS, an optical imager that will reach
an AB magnitude depth of 24.5 \citep[for extended sources at $10-\sigma$, see
also][]{Scaramella2021} with a single broad
$I_{\scriptscriptstyle{\rm E}}$ filter, and NISP, a combined
near-infrared imager \citep[in $Y_{\scriptscriptstyle{\rm E}}$,
$J_{\scriptscriptstyle{\rm E}}$, and $H_{\scriptscriptstyle{\rm E}}$,
see][]{Schirmer2022} and slitless spectrograph. The optical
imager will determine galaxy shape distortions with unprecedented
accuracy. When combined with a precise determination of the true
ensemble redshift distribution, this allows  the weak
lensing effects caused by the distribution of matter along the line of
sight to be measured and the  cosmological parameters to be constrained
\citep{Blanchard2020}.

The estimated number of weak lensing source galaxies that will
be imaged from {\it Euclid} makes their systematic spectroscopic
follow-up unfeasible; this mission is thus critically dependent upon
the determination of accurate photometric redshifts ($z_{\rm
  phot}$). Currently, the precision of photometric redshifts
based on multi-band optical surveys is to the order of $\sigma_z/(1+z)
= 0.03-0.06,$ and the fraction of catastrophic outliers, which are  defined as
objects whose $z_{\rm phot}$ differs from their spectroscopic redshift
($z_{\rm spec}$) by more than $0.15(1+z),$ is of the order of a few
tens of percent \citep{Ma2006,Hildebrandt2010}. We expect that the
combination of ground-based optical and {\it Euclid} near-infrared photometry
will deliver the slightly improved requirements of the mission: 
$\sigma_z/(1+z)\le 0.05,$ and a fraction of catastrophic outliers less
than 10\% (see \citealt{Laureijs2011} and \citealt{Desprez2020}).

While small changes in $z_{\rm phot}$ precision per source have a
relatively small impact on cosmological parameter estimates, small
systematic errors in $z_{\rm phot}$ can dominate all other
uncertainties for these experiments. The aim of the  C3R2 project
\citep{Masters2015,Masters2017,Masters2019,Guglielmo2020,
  Stanford2021} is to  calibrate photometric redshifts by measuring
accurate spectroscopic redshifts of selected objects sampling a
self-organizing map (SOM), a representation of the galaxy
color space. The SOM projects the high-dimensional galaxy color
space onto a 2D plane. Each galaxy is assigned to a cell in this plane
with given coordinates $(X,Y)$. If this plane is sampled with enough
cells, the distribution of photometric redshifts of all galaxies
belonging to a cell is narrow and there is a well-defined
correspondence between the position occupied by a galaxy in the
multi-color space and its redshift. We can define $z_{\rm phot,SOM}$
as the average of the photometric redshifts of all galaxies belonging
to the cell. By measuring spectroscopic redshifts of galaxies in each
cell it is possible to calibrate the mean of the photometric redshift
distribution in an efficient and homogeneous way across the galaxy
color space. Regions in color space particularly difficult to
calibrate because of broad or bimodal photometric redshift
distributions can be identified by comparing the measured
spectroscopic redshifts with $z_{\rm phot,SOM}$ and looking for
deviations larger than $0.15(1+z)$. The minimum calibration
requirement is to populate each SOM cell with at least one
spectroscopic redshift; problematic cells might require (much) more
than this. Ultimately, galaxies belonging to uncalibrated cells may be
dropped from the {\it Euclid} weak lensing sample; for a detailed
discussion see \citet{Masters2015}.

In this work we continue this effort by presenting the redshift
measurements of $z>1$ galaxies collected at the Large Binocular
Telescope (LBT) in the COSMOS \citep{Capak2007,Scoville2007,Lilly2007}
and VVDS \citep{McCracken2003,LeFevre2004,Jarvis2013} fields, using
the near-infrared LUCI spectrographs. Further data releases of optical
spectra acquired at the VLT and the GRANTECAN telescopes are in
preparation. The {\it Euclid} science working groups are actively discussing
how the final ground-based dataset will be merged with the
spectroscopy information delivered by the {\it Euclid} mission itself to
perform the optimal calibration of the photometric redshifts.

The paper is organized as follows: in Sect. \ref{Sec:Masks} we describe the
strategy, target selection, and mask preparation; in
Sect. \ref{Sec:observations_data_reduction} we describe the
observations and data reduction; in
Sect. \ref{sec:redshift_assignment} we discuss the redshift
determination and the attribution of a flagging scheme consistent over
the whole C3R2 survey; in Sect. \ref{Sec:results} we present the
results of the redshift assignments, and we investigate the bias of the
photometric redshifts using in C3R2 and the origin of catastrophic SOM
redshifts, and we discuss our success rate and SOM cell
coverage; finally, we present our conclusions in
Sect. \ref{Sec:conclusions}.


\begin{figure*}[h!]
    \centering
    \includegraphics[scale=0.45]{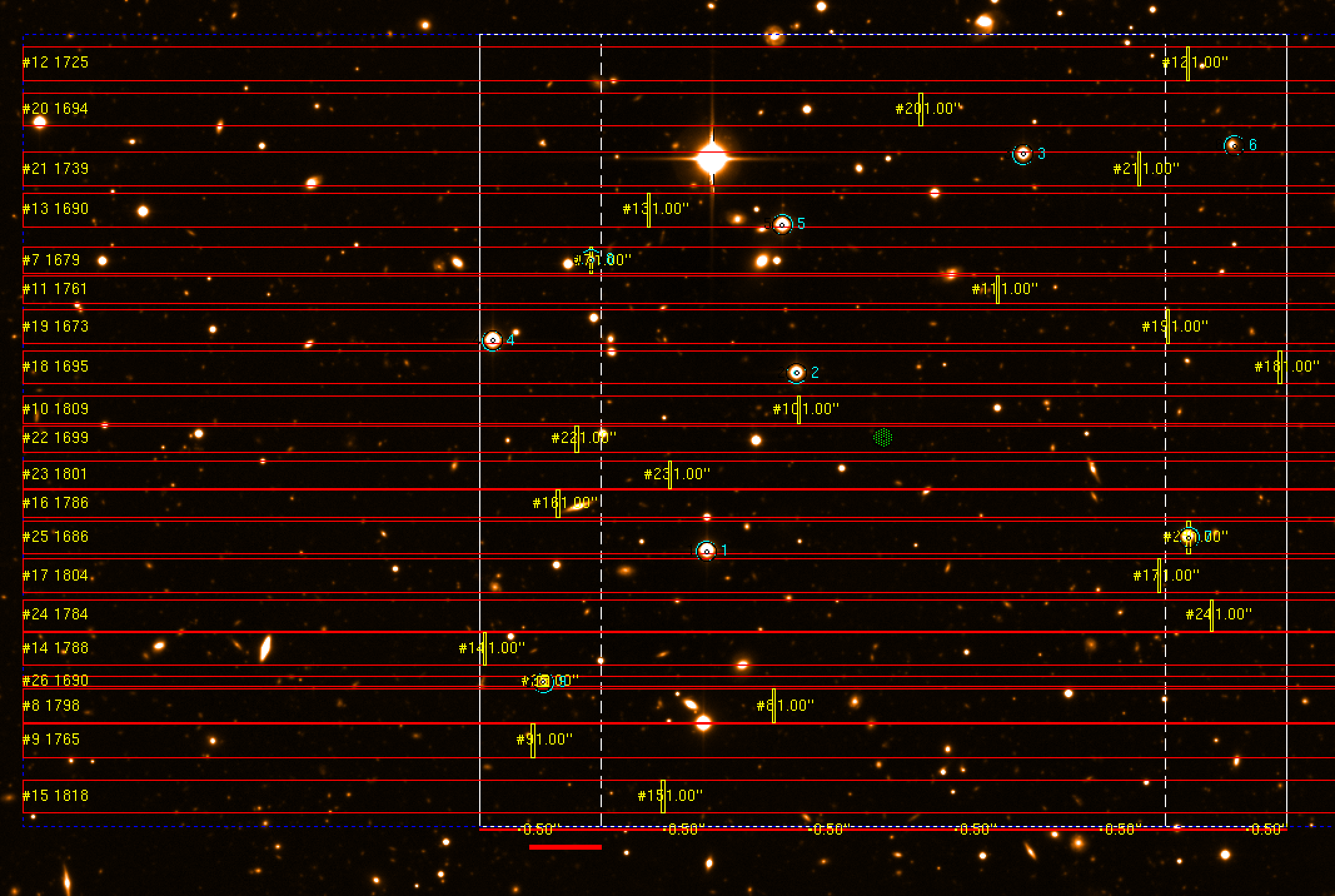}
    \caption{COSMOS\_M25H mask designed with the {\tt lms}
      tool. The white square shows the LUCI field of view $(4\times 4
      \, {\rm arcmin} ^2)$,
    the long-dashed vertical lines bracket the optimal field of view
    for spectroscopy (2.8\,arcmin wide). The green circle at the center of the field allows  the mask to be moved. The cyan circles identify the
    alignment stars. The yellow rectangles show the slits; slit number 25
    is positioned on an acquisition star to monitor seeing and
    transparency during the observations. The square slit
    number 26, also positioned  on an acquisition star, allows 
    verification of the centering of the mask after alignment. At the lower
  end of the mask, a series of six small holes is present for
  engineering purposes. Below these, the red rectangle is the area occupied by the identification number cut into the mask.}
    \label{fig:masksH}
\end{figure*}

\begin{figure*}
    \centering
        \includegraphics[scale=0.45]{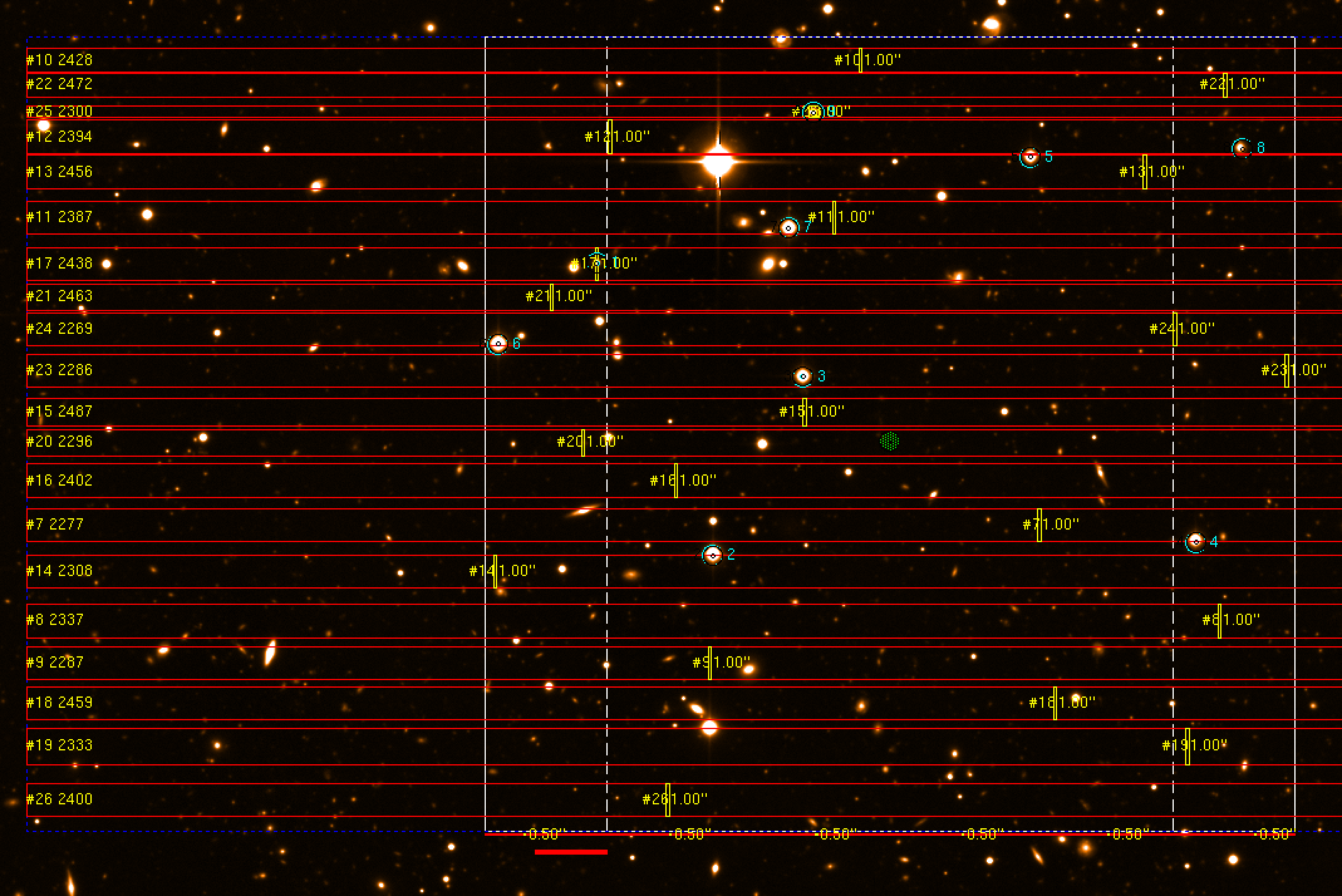}
    \caption{COSMOS\_M25K mask (see       Fig. \ref{fig:masksH} for a description). Most of the
      slits are assigned to different objects.}
    \label{fig:masksK}
\end{figure*}

\section{Strategy, target selection, and mask preparation}
\label{Sec:Masks}

The LBT consists of two 8\,m mirrors mounted on a common structure and
pointing at the same position on the sky. The LUCI1 and LUCI2
near-infrared spectrographs \citep{Seifert2003} are mounted at the
front Bent Gregorian f/15 focal stations of the LBT, and can be used
in combination with masks \citep{Buschkamp2010}, designed and cut well
in advance of the observations, that allow the simultaneous collection
of multiple spectra. The field must be observed with identical
pointings and rotation angles, but potentially with different masks
and instrumental setups on the two sides.  The masks cover a $4\times
4\, {\rm arcmin}^2$ field of view, but only for slits placed in a
2.8\,arcmin wide central stripe are optimally focused spectra
delivered. Moreover, the wavelength range covered depends on the
distance of a slit from the middle of a mask. At least three stars are
needed to align the masks. Dedicated holes (typically $4\times 4\,{\rm
  arcsec}^2$ big) were required until the June 2020 runs. Starting
from the July 2020 run, after a change in requirements from the
observatory, only one hole was necessary.

\subsection{Strategy}

Given the characteristics of the LBT and of the LUCI spectrographs described above, and
following the strategy adopted by \citet{Guglielmo2020} (hereafter
G2020), we collected spectra in  the $H$ and $K$ bands. In the $H$ band
we can measure spectroscopic redshifts of galaxies between 1.3 and 1.7
by detecting their H$\alpha$, [\ion{N}{ii}], and [\ion{S}{ii}] emission
lines, or of galaxies with redshifts between 2 and 2.7 by detecting
their H$\beta$ and [\ion{O}{iii}] lines. In the $K$ band we can
measure spectroscopic redshifts of galaxies between 2 and 2.7 by
detecting their H$\alpha$, [\ion{N}{ii}], and [\ion{S}{ii}] emission
lines, or (in principle) of galaxies with photometric redshifts less
than 1.7 by detecting their Paschen (hereafter Pa) lines.
We used the N1.8 camera,
which delivers $0.25\times0.25\,{\rm arcsec}^2$ per pixels, and the G210
gratings, for which 1\,arcsec-wide slits give a resolution of $R=2950$ in
the $H$ band and $R=2500$ in the $K$ band. The nominal wavelength
range for a centered slit is 0.202\,$\mu$m and 0.328\,$\mu$m in the
$H$ and $K$ band, respectively. The central wavelength can be
adjusted in the range 1.55 to 1.75\,$\mu$m for the $H$ band, and 2.06
to 2.40\,$\mu$m for the $K$ band. We optimized the central wavelength
for each mask separately, based on the photometric redshifts of the
observed galaxies, to maximize the likely return.

The typical angular size of the galaxies in the C3R2 catalogues is 1--2\,arcsec in diameter. In order to achieve maximum efficiency avoiding separate
sky observations, we opted for an on-slit nodding strategy, which sets
the default slit length to 10\,arcsec and forces all slits to be
parallel. Taking into account the minimum allowed distance between
slits, and the necessity of cutting holes for acquisition stars, the
maximum number of galaxies targeted per mask is $\approx 20$. In reality, we
never managed to observe more than 18 galaxies per mask.
We ranked the list of positions and rotation angles
according to the total number of galaxies observable simultaneously in
the $H$ and $K$ bands. We observed the 
VVDS field in October and the COSMOS field in the months from December to May.
  We ended up with a list of 13 pairs of
($H+K$) masks in the VVDS field, and 28 pairs of ($H+K$) masks in the
COSMOS field, for which at least a total ($H+K$) 20 galaxies could be
assigned, minimizing the number of repeated observations. As in G2020,
we aimed to observe each mask for 2 hours, split into 36 exposures of
200\,s each, dithering the mask along the slits by 2\,arcsec following
an ABBA pattern. The actual number of exposures collected for each
mask varied according to observational conditions and constraints; see
Table \ref{Tab:observed_pointings}.

\subsection{Target selection and mask preparation}

The selection of galaxies to be observed started from the catalogues
produced by \citet{Masters2019} and used by G2020, excluding galaxies
already observed with KMOS at the VLT. The catalogues adopt the
photometric redshifts provided by \citet{Ilbert2006} and
\citet{Laigle2016}.  We considered Priority 1 and Priority 2 galaxies
(hereafter {\it primary} galaxies), as in G2020, and
mapped the number of galaxies assignable to each mask as a function of
the coordinates of the field centers and rotation angles, taking into
consideration several constraints. At least one suitable guiding star
had to be present in the allowed patrol field and within the
appropriate magnitude range; at least three acquisition stars
(preferably selected to have low proper motions) within the
appropriate magnitude range should be present in the $4\times4\,{\rm
  arcmin}^2$ field of view; at least three  $4\times 4\, {\rm arcsec}^2$ holes (or just one from October
2020) were cut for this purpose. At
least one $10 \times 1\,{\rm arcsec}^2$ slit was assigned to a star,
to be able to reconstruct empirically the dithering pattern and
measure the seeing and the relative transparency during the
observations. The galaxies were placed such that the expected
H$\alpha$ line based on their photometric redshift fell into the
wavelength range computed based on the distance to the central stripe
of the field and the optimized central wavelength.

The actual design of the masks was performed with the {\tt lms}
tool\footnote{https://sites.google.com/a/lbto.org/luci/preparing-to-observe/mask-preparation/lms-install}.
During this phase, further tweakings of the centers, position
angles, and  slit assignments were necessary. Guide stars too
near the border of the patrol field had to be changed,
some acquisition stars or some slits
had to be dropped or changed in length
(down to 7\,arcsec) because of additional constraints (e.g., one end of the
mask has a regular pattern of small holes that must be
avoided). Finally, the space available between the slits was filled
manually with  {\it secondary} galaxies, when possible. These are
galaxies where the [\ion{O}{iii}] lines could possibly appear in the $H$ band,
or the Pa lines in the $K$ band. Once the list of
Priority 1 and 2 galaxies assigned to each mask was ready, the optimal
central wavelength was calculated as the average of the H$\alpha$
wavelengths, redshifted with the respective photometric redshifts.
The {\it gbr} files detailing the masks for the cutting machine were
passed to the observatory, where the masks were produced and inserted
into the cryostatic dewar before each observing run.

An example of a pair of $H+K$ masks is given in the
Figs. \ref{fig:masksH} and \ref{fig:masksK}.  All the observed masks
are listed in Table \ref{Tab:observed_pointings}, where the
coordinates of their centers and of their rotation angles, together
with the date of the observations are given.  In the end (see Table
\ref{Tab:observed_pointings}), on average we extracted 12.3 galaxy
spectra per mask, of which on average 9.6 were {\it primary}.

\section{Observations and data reduction}
\label{Sec:observations_data_reduction}

The scripts controlling the telescope and LUCI operations were
prepared using the LBTO OT
software\footnote{https://sites.google.com/a/lbto.org/observing-tool-manual/observing-tool/ot-installation}. Except
when one of the two LUCI spectrographs was not available, the paired
approach was adopted, allowing simultaneous observation of the same
field in the $H$ and $K$ bands. Monocular scripts were prepared for
the October 2019 run (when only LUCI1 was available) and for the
January 2020 run (when LUCI1 became unavailable during the
run). Monocular scripts were also prepared for the observations of
telluric standards. Calibration scripts to obtain dark, flat, and arc
observations were also prepared in paired mode and performed
during the day or at night during periods of bad weather.
The observations started in visitor mode (October 2019), and continued
in remote observing mode as the COVID19 emergency made international
traveling impossible. Staff members from the LBT Observatory operated
the instrument from Tucson for all observing runs, with scientists
from the Max Planck Institute for extraterrestrial physics (M. Fabricius, S. de Nicola, R. Saglia, J. Snigula) and
the Landessternwarte
(J. Heidt) mainly supervising, but also directly controlling the
procedures from Germany. Overall, the VVDS field was observed under good
meteorological conditions; in contrast, several nights during which the
COSMOS field was observed  (in particular the March 2021 run), suffered
from cirrus, high winds, and poor seeing.

The observation of a field follows four steps. First, an image is taken
without the mask at each telescope, the acquisition stars are
identified, and their positions are measured. Second,  the best-fitting field
rotations and translations are determined separately for the two
images and applied, possibly culling the most deviant acquisition
stars. Third,  a second pair of images is taken through the mask to verify
that acquisition and seeing-monitoring stars appear in the appropriate
holes and slits.  If necessary, a second translation is determined and
applied to optimize the centering orthogonal to the slits.  The
achieved RMS precision of the alignment was typically between 0.1 and
0.3\,arcsec. Fourth, the spectroscopic observations start.

After the observation of a field, or before the observation of the
next one, a telluric standard was observed separately for the $H$ and
$K$ band, putting the standard star sequentially in three slits of each
mask, selected to cover the whole probed wavelength range. We observed
a total of 88 masks (58 in the COSMOS field and 30 in the VVDS field),
47 in the $H$ band (30 in the COSMOS field and 17 in the VVDS field), and
41 in the $K$ band (28 in the COSMOS field and 13 in the VVDS field). As
mentioned above, during the October 2019 run, LUCI2 was unavailable,
and during the January 2020 run  LUCI1 stopped working.

The data reduction was performed using the {\tt IDL Flame} pipeline
developed by \citet{Belli2018}. We used the configuration which relies on the
science data for tracing, extraction, and wavelength calibration,
exploiting the brightness of the sky background.  The steps follow the
sequence described in \citet{Belli2018}. First, the position of the
reference star on each frame is searched for and, if detected, the flux,
vertical position, and full width at half maximum (FWHM) are measured. This allows the determination
of the nodding and dithering offsets of each A and B frame in the ABBA
sequences. However, for many of the $K$-band pointings, the reference
star cannot be detected in single frames and appears visible only
after the coaddition is performed. In these cases, the list of the
nominal shifts is used. After application of the bad pixel and master
pixel maps, a master slit flat is computed and used to identify the
slit positions and map their edges.  Cutouts of the slits are produced
and, after a rough wavelength calibration, individual OH emission
lines are identified and fitted with a Gaussian. The relation between
pixel position and wavelength is determined as a second-order polynomial
for each slit row. The spatial illumination correction is determined
from the detected OH lines and applied. A model of the sky is
constructed following \citet{Kelson2003} and subtracted. Finally, each
sky-subtracted slit frame is rectified and wavelength calibrated
before all the A and B frames are stacked together and the A-B and B-A
results are produced. Their combination gives the final 2D spectra
shown in Figs. \ref{fig:spec1d2dH} and
\ref{fig:spec1d2dK}. One-dimensional spectra are extracted after
having identified the appropriate spatial window, either determining
the spatial extent of the emission lines or of the continuum. Examples are
shown in  Figs. \ref{fig:spec1d2dH} and \ref{fig:spec1d2dK}. We note
that wavelengths refer to the vacuum and do not take into account the
barycentric correction, which is applied a posteriori to the measured
redshift.


\section{Redshift determination and flagging}
\label{sec:redshift_assignment}

Redshifts are measured on the 1D spectra after having spotted by eye the
signatures of emission lines on the 2D spectra:  the presence of a
sequence of a
negative (black), a positive (white), and a negative (black) blob (see
Figs. \ref{fig:spec1d2dH} and \ref{fig:spec1d2dK}). Based
on the alignment of the predicted positions of H$\alpha$, [\ion{N}{ii}], [\ion{S}{ii}],
H$\beta$, [\ion{O}{iii}], or Pa lines with the peaks of the observed emission
lines, an estimate of the redshift of the galaxy is derived, together
with the appropriate flag. Following the scheme used in
G2020, a ${\rm Flag}=4$ is assigned to redshifts with
multiple good S/N line detections; a  ${\rm Flag}=3.5$ is given to very good
S/N single line detections; a  ${\rm Flag}=3$ is given to convincing single
line detections. Lower ${\rm Flag\,} (1 {\rm \,and\,} 2)$ spectra indicate
redshifts that
are not secure enough for our purposes;   ${\rm Flag}=-99$ are non-detections
or failed reductions. Examples of  ${\rm Flag}=4, 3.5, 3$ spectra and redshifts
are given in  Figs. \ref{fig:spec1d2dH} and \ref{fig:spec1d2dK}.

In a number of slits, more than
one object was detected, either as a spatially separate object or on
the same line of sight, but with different redshifts. In these cases
multiple 1D spectra were extracted and analyzed. The correct
identification was attempted by inspection of the relevant images.

The procedure described above was performed independently by two of us
(RPS and RZ); discrepant assessments were discussed and resolved. We
extracted 1119 spectra, of which 19 were not matched with an object
from the photometric parent catalogues.  We matched 292 good spectra (${\rm
  Flag}\ge 3$): 163 in the COSMOS field, 129 in the VVDS
field. The complete statistics are given in Table
\ref{Tab:spectra_statistics}, where we list the numbers referring to
    {\it primary} spectra (targeting the H$\alpha$ line) and {\it
      secondary} spectra (targeting the [\ion{O}{iii}] or Pa lines).


\begin{table*}
    \caption{Statistics of observed spectra.}
    \label{Tab:spectra_statistics}
  \begin{tabular}{lccccccccc}
    \hline
    Spectra   & All & $H$ band & $K$ band & COSMOS & $H$ band & $K$ band & VVDS    & $H$ band & $K$ band \\
  \hline
Extracted     &    1119  &   635    & 484  &  766       & 424   &      342   &353 &   211    &  142 \\
Good            &      305  &  212     &   93  &  174      & 127    &      47   & 131  &    85    &   46 \\
Identified     &    1100   &  618   &  482   & 751      &  410   &    341 &  349  &  208    &  141 \\
Good            &      292   &  200    &   92  &  163      &  117    &    46 &   129  &    83    &    46 \\
Primary         &      851   &  446   &  405  &  597      &  316   &  281  & 254  &  130   &  124     \\
Good             &    249    &  158    &  91  &  139       &     94  &       45   & 110   &    64    &   46 \\
Secondary     &     249    & 172   &    77  &   154     &      94  &      60 &    95   &      78   &   17 \\
Good             &       43     &  42   &      1  &    24      &      23  &      1 &    19    &     19    &     0 \\
  \hline
Unidentified  &     19     & 17      &    2   &    16       &     15    &     1 & 4  &     3   &        1 \\
Good             &     13     & 12      &    1    &   11        &    10   &      1 & 2   &    2   &        0 \\
Primary         &     14     & 12      &    2     &  11         &    10   &     1 & 3   &   2   &        1 \\
Good             &     8      &   7   &       1     &    7        &      6     &   1   & 1   &  1   &       0 \\
Secondary     &     5       & 5     &     0      &    4        &      4    & 0  &  1     &  1   &        0 \\
Good             &     5      & 5 &         0    &      4      &      4       &  0&   1   &    1  &         0 \\
 \hline
  \end{tabular}
\end{table*}

Table \ref{Tab:observed_pointings} (column ``Success Rate'') gives the
number of extracted identified spectra (a total of 1100) in each
pointing together with the number of good spectra $( {\rm Flag}\ge
3)$ and the number of primary spectra collected.  The mean success
rate (defined as the ratio of the number of identified good spectra
over the total number of collected identified spectra) is 0.27 (0.22
in the COSMOS field and 0.37 in the VVDS field; the lower success rate
in the COSMOS field stems from the suboptimal meteorological
conditions under which several COSMOS masks were observed).  It is
higher in the $H$ band (0.32; 0.28 in the COSMOS field and 0.40 in the
VVDS field) than in the $K$ band (0.19; 0.13 in the COSMOS field and
0.33 in the VVDS field). The success rates achieved by G2020 are
better, which  stems from the larger field of view (which allowed an
optimized choice of good candidates), the fixed wavelength range, and
the IFU available with the KMOS instrument. We investigate the
properties of the galaxies with unreliable spectra $({\rm Flag}< 3)$ 
in Sect. \ref{Sec:results}.

 The barycentric corrections
$z_{\rm helcorr}=v_{\rm helcorr}/c$ appropriate to each
mask are listed in Table \ref{Tab:observed_pointings} together with
the Julian date (JD) adopted for the computation. This  corresponds to the
middle of the sequence of exposures. When two series of exposures observed in
different nights are summed together, the JD refers to the first of the two.
The barycentric corrections are
computed through the {\tt python} routine  {\tt pyals.helcorr}  from
{\tt PyAstronomy}; we quote the two digits that 
affect the last quoted digit of the redshifts. Using the
relation $1+z_{\rm spec} =(1+z_{\rm meas} ) \, (1+z_{\rm helcorr})$, where
 $z_{\rm meas}$ are the measured redshifts shown in
 Figs. \ref{fig:spec1d2dH} and \ref{fig:spec1d2dK}, we
 compute the redshifts $z_{\rm spec}$ given in Table \ref{Tab:redshifts}
 using the formula $z_{\rm spec}=z_{\rm meas} +(1+z_{\rm meas} )  \, 
 z_{\rm helcorr}$. In addition, Table \ref{Tab:redshifts} reports the object ID, its
 coordinates, the Flag, the line or ensemble of lines used (H$\alpha$ plus
 possibly [\ion{N}{ii}] and/or  [\ion{S}{ii}], [\ion{O}{iii}] plus possibly H$\beta$, Pa) and the
 name of the spectrum.

\begin{figure*}
    \centering
\includegraphics[scale=.29]{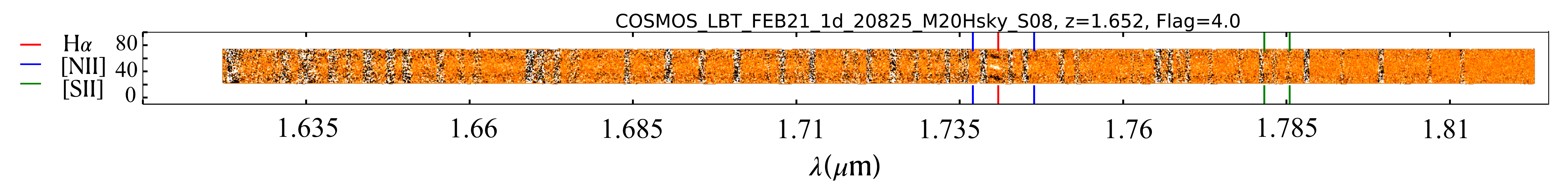}  
\includegraphics[scale=.9]{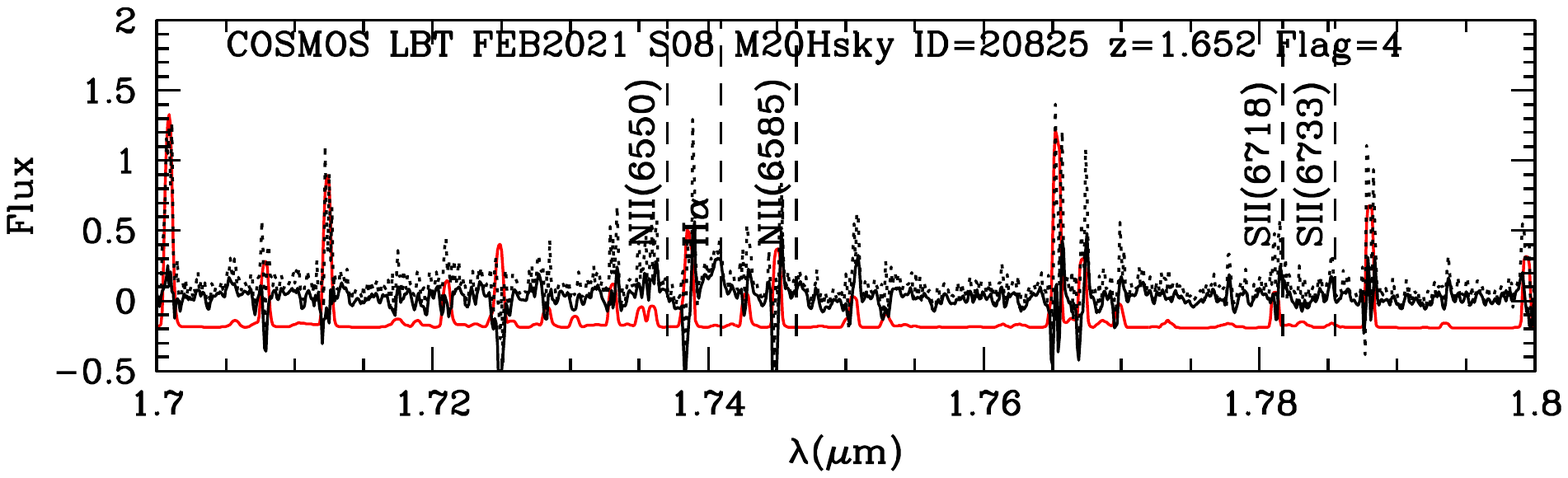}  %

\includegraphics[scale=.29]{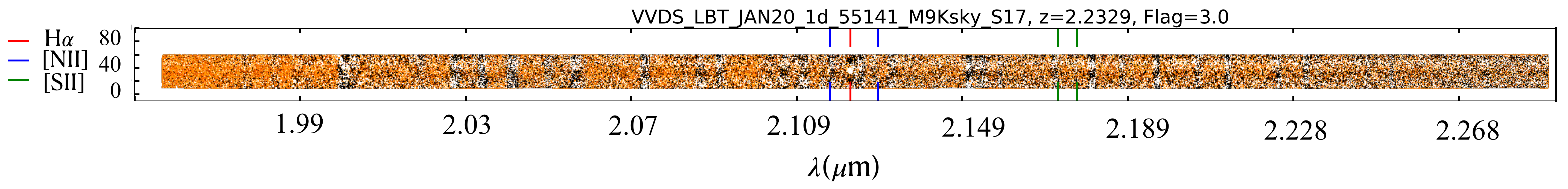}     
    \includegraphics[scale=.9]{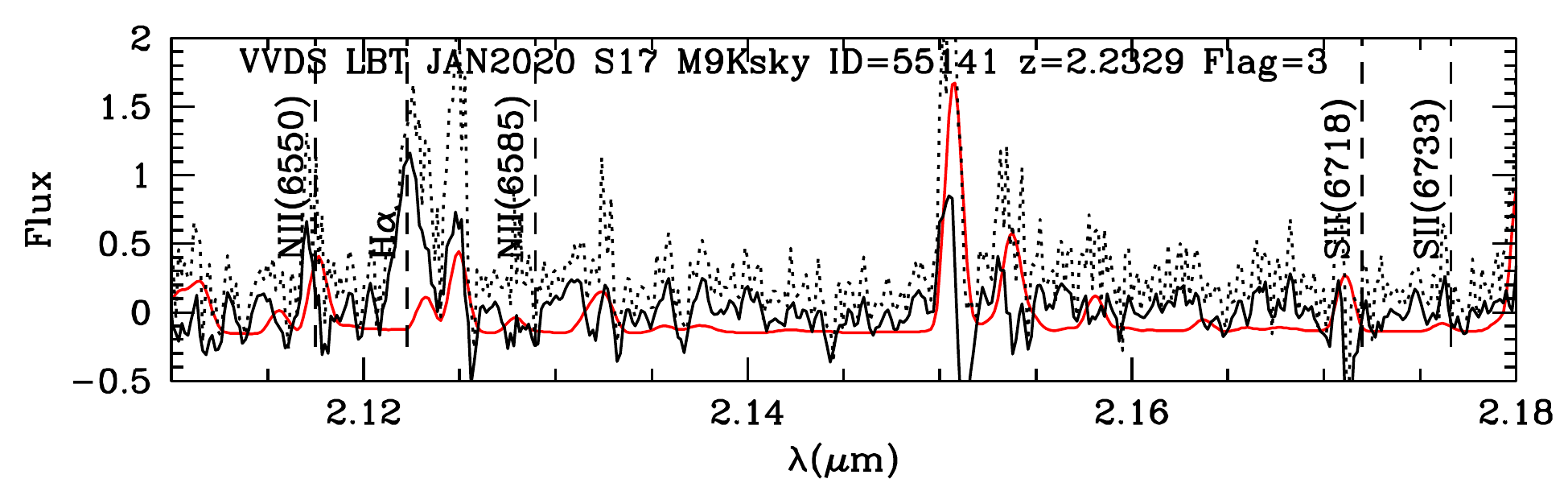} 
     \caption{One- and two-dimensional spectra with ${\rm Flag}=4,
       3.5, 3$ of objects with $z_{\rm meas}<2$. The 2D 
       spectra are 10\,arcsec in width, with the vertical scale given in
       pixels, and show the whole wavelength range (in $\mu$m)
       observed. The 1D  spectra show the wavelength range
       around the relevant emission lines; the black solid lines show
       the measured flux (in arbitrary units with a 3 pixel
       smoothing), the dotted lines the flux without smoothing plus
       its errors, the red lines the sky (scaled down by a factor of
       1000 and shifted by $-0.2$ flux units).  The vertical dashed
       lines indicate the emission lines (with vacuum rest-frame
       wavelengths in \AA ) used (when visible) to measure the
       spectroscopic redshift (not yet corrected to the heliocentric
       system).}
    \label{fig:spec1d2dH}
  \end{figure*}

\setcounter{figure}{2}
  \begin{figure*}
  \centering
  \includegraphics[scale=.285]{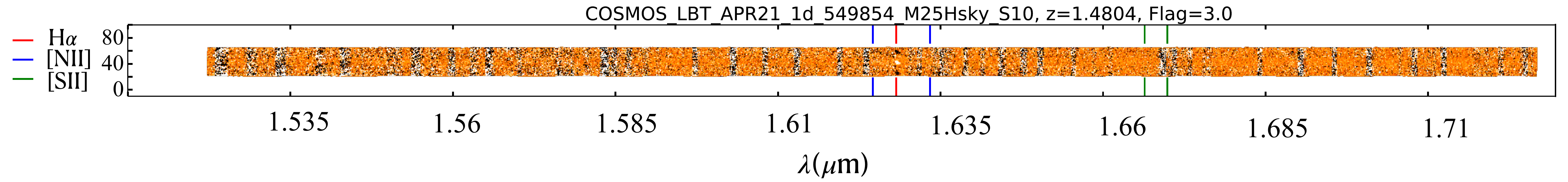}    
     \includegraphics[scale=.9]{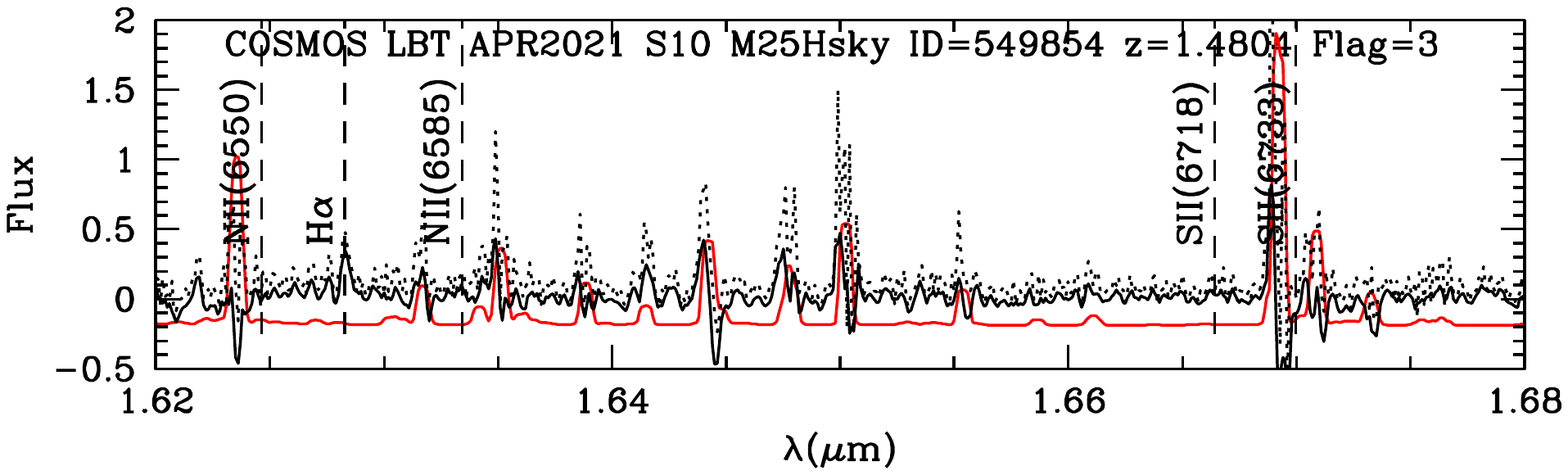}
     \caption{continued. }
\end{figure*}

\begin{figure*}
    \centering
          \includegraphics[scale=.29]{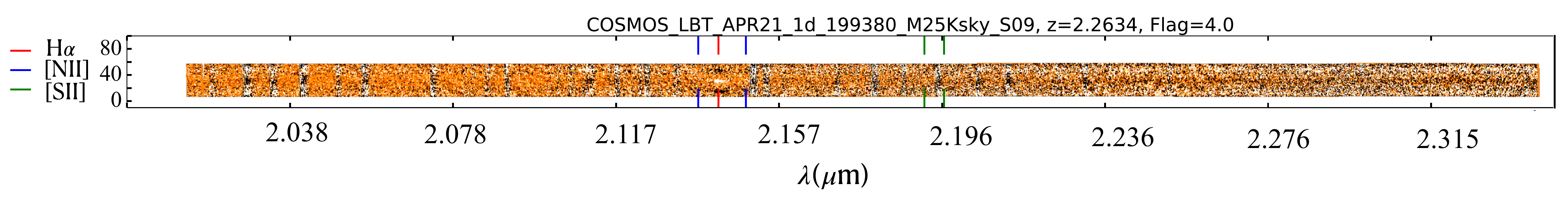}     
     \includegraphics[scale=.9]{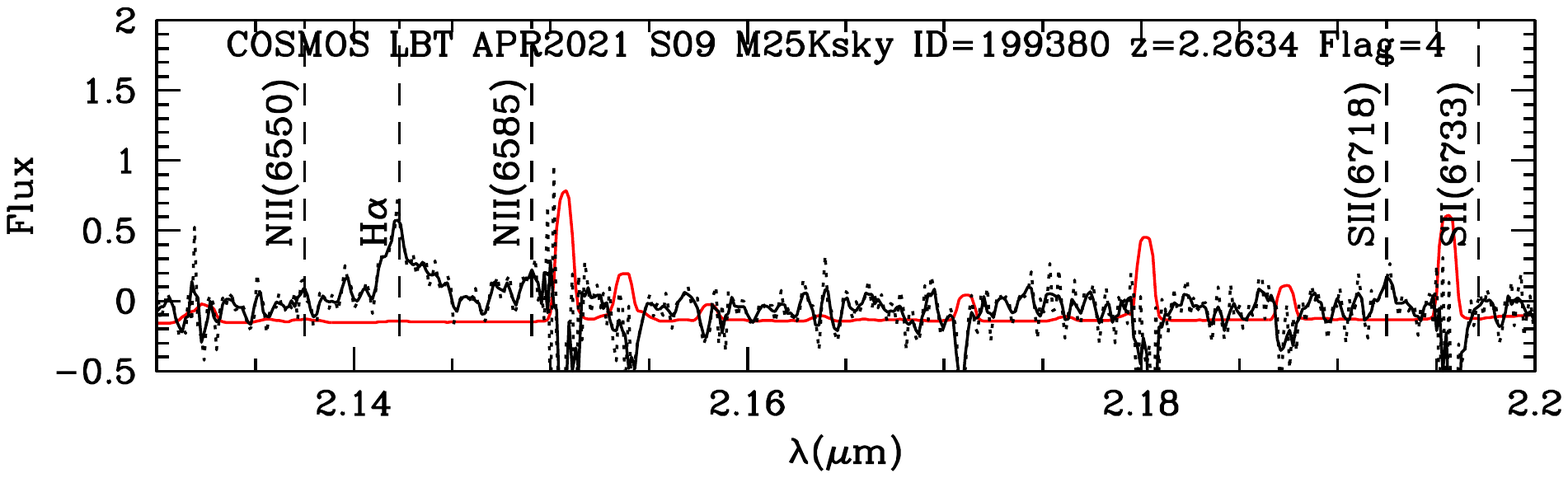} 
 \includegraphics[scale=.3]{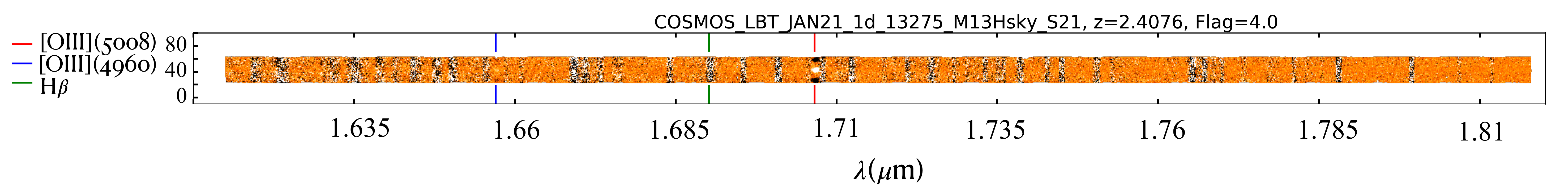}  
     \includegraphics[scale=.9]{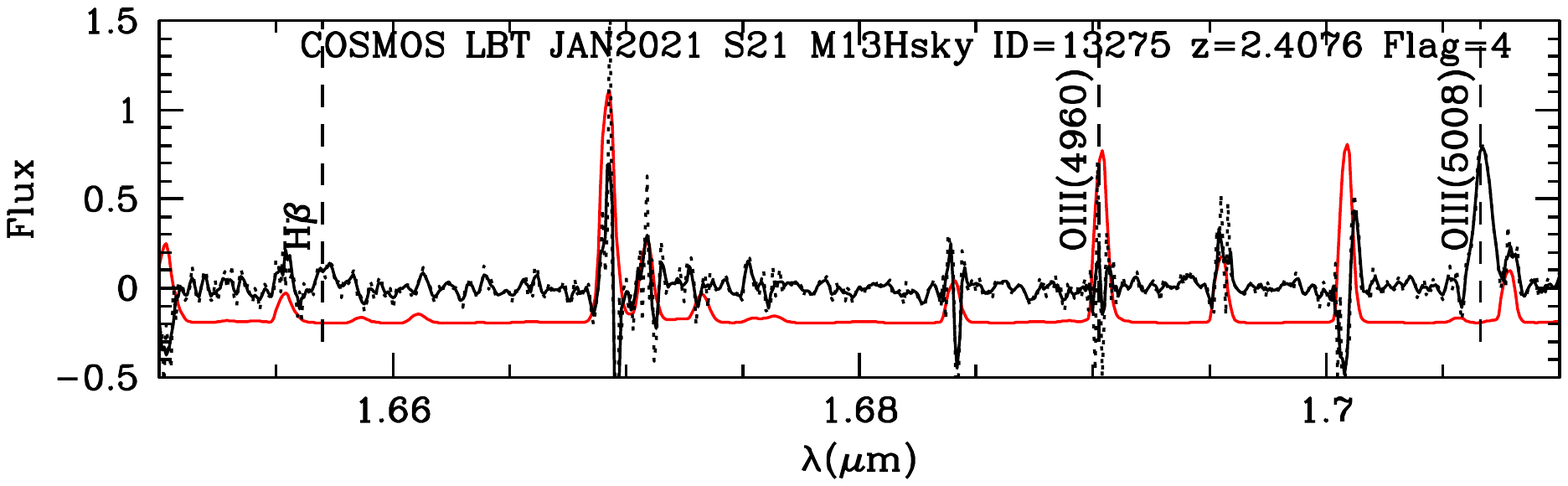}
\includegraphics[scale=.29]{cVVDS55141_2D.pdf} 
     \includegraphics[scale=.9]{cVVDS_LBT_JAN20_1d_55141_M9Ksky_S17.pdf} 
     \caption{One- and two-dimensional spectra with ${\rm Flag}=4$ and 3  of
  objects with $z_{\rm meas}>2$. Lines as in Fig. \ref{fig:spec1d2dH}.}
    \label{fig:spec1d2dK}
\end{figure*}

\begin{figure}[h!]
    \centering
    \includegraphics[scale=0.45]{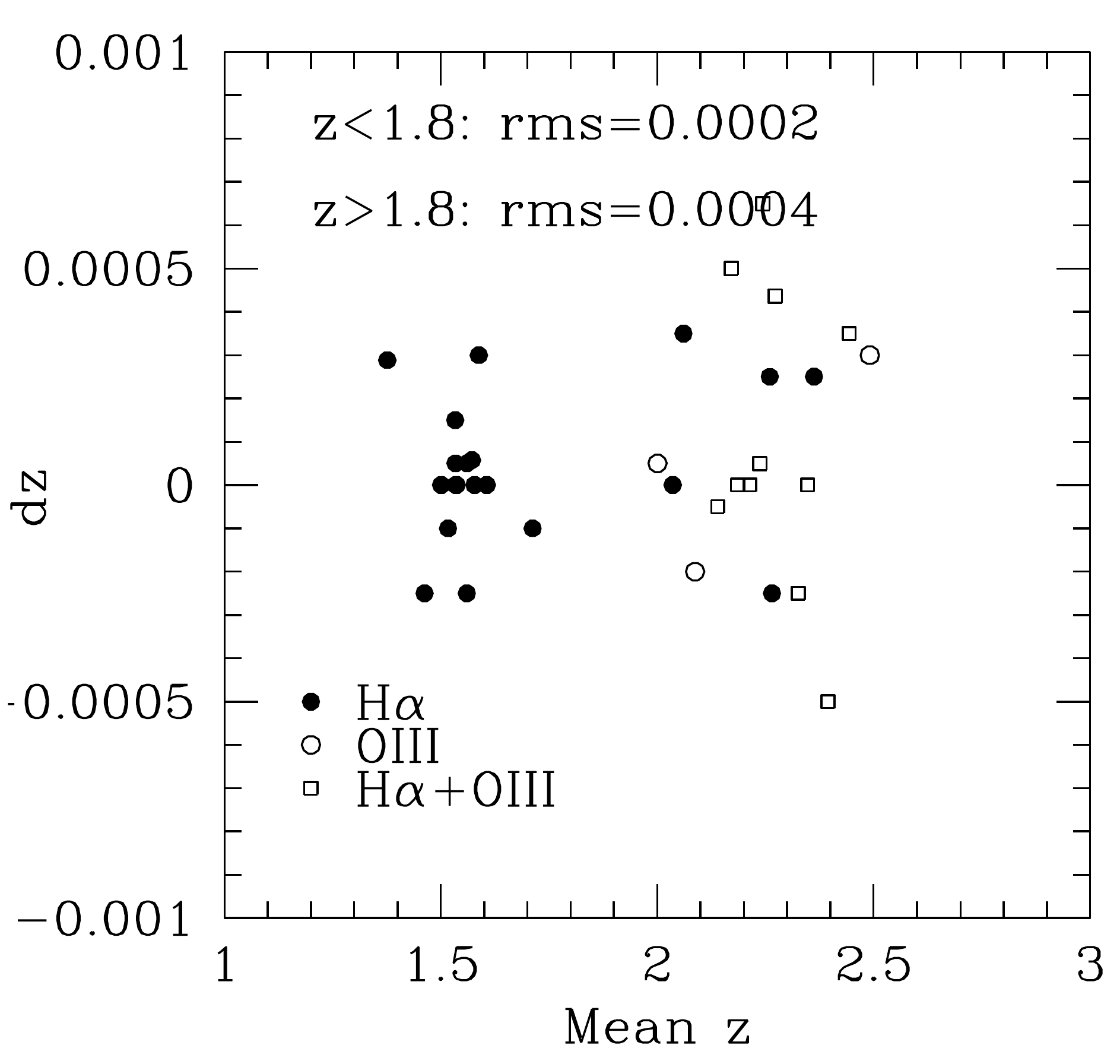}
    \caption{Errors from repeated observations.
      The filled circles show the differences
      d$z$ between the spectroscopic redshifts determined from two or
      more spectra of the same objects using the H$\alpha$ line. If
      more than two spectra were collected, the difference to
      the average of the measurements is shown. The open circles show
      differences between redshifts based on the [\ion{O}{iii}] lines; the
      open squares
      show differences between redshifts measured using the
    H$\alpha$ and the  [\ion{O}{iii}] lines.}
    \label{fig:errors}
\end{figure}

A number of objects were observed multiple times, enabling an
empirical determination of the errors on our redshifts and to compare
redshifts derived from the  $K$- and $H$-band spectra. Figure
\ref{fig:errors} shows that the RMS uncertainty on the redshifts of objects
at $z<1.8$ with repeat observations is
0.0002, and two times larger for objects with $z>1.8$. One COSMOS object
$({\rm ID}=481647)$ was observed by the previous C3R2 releases, providing
$z_{\rm spec}=1.501$,  identical to $z_{\rm spec}=1.501$ obtained here. A
further COSMOS object $({\rm ID}=480046)$ was observed by
\citet{Stanford2021}, who reported $z_{\rm spec}=2.2705$, while our
measurement is  $z_{\rm spec}=2.2702$, in line with the results discussed above.
Following these
findings, we quote the measured redshifts to four decimal digits. The redshifts
determined from H$\alpha$ and  [\ion{O}{iii}] lines agree within the errors,
with no appreciable bias.



A unique redshift $\langle z \rangle$ and flag were assigned to each object with repeated
observations by averaging the available measurements; the resulting
values are reported in Table  \ref{Tab:meanrepeats} and used in the
following figures.

\begin{table*}
  \caption{Mean redshift and flag for objects with multiple spectra.}
  \label{Tab:meanrepeats}
 \begin{tabular} {lrccccc}
  \hline
Field & OBJ\_ID & $\langle z \rangle$ & $|\Delta z|/2$ or RMS & N &
                                                               $\langle {\rm Flag} \rangle$ & Lines\\
 \hline
 COSMOS  & 32961    &   1.4623  & 0.0003  & 2 & 3.75   &  H$\alpha$  \\
 COSMOS &  38150   &   2.3945  &  0.0005 & 2 &  3.5  &  H$\alpha$+[\ion{O}{iii}]  \\
 COSMOS & 402635  &   2.1708  &  0.0005 & 2 &  3.5  &  H$\alpha$+[\ion{O}{iii}]  \\
 COSMOS & 462025  &   2.0875  & 0.0002  & 2 &  3     &   [\ion{O}{iii}]  \\
 COSMOS & 463893  &   2.3473  &  0          & 2 &  3.5  &  H$\alpha$+[\ion{O}{iii}]  \\
 COSMOS & 472716  &   1.7119  & 0.0001  & 2 &  3.25  &  H$\alpha$  \\
 COSMOS  & 479007  &   2.2725  & 0.0004  & 3 &  4      &  H$\alpha$+[\ion{O}{iii}]  \\
 COSMOS & 481315  &   2.4910  & 0.0003  & 2 &  4       &   [\ion{O}{iii}]  \\
 COSMOS  & 481647  &   1.5010  & 0           & 2 & 4        &  H$\alpha$  \\
 COSMOS & 484310  &   1.5879  & 0.0003  & 2 &  3.25  &  H$\alpha$  \\
 COSMOS & 811438  &   2.3620  & 0.0003  & 2 &  3       &  H$\alpha$  \\
 VVDS      & 3324      &   1.5340   & 0.0001   & 2 &  4         &  H$\alpha$  \\
 VVDS      & 7941      &   1.5346   & 0   & 2 &  3.25     &  H$\alpha$  \\
 VVDS      & 22805     &   1.5333   & 0.0001  & 2 &  3.25   &  H$\alpha$  \\
 VVDS      & 24882    &   1.3763   &  0.0003 & 3 & 3.333 &   H$\alpha$  \\
 VVDS      & 24903     &   1.5781   & 0   & 2 &   3        &  H$\alpha$  \\
 VVDS      & 36412     &   1.5366   & 0   & 2 &   4         &  H$\alpha$  \\
 VVDS      & 37921     &   2.0007   & 0.0001   & 2 &   3.5    &   [\ion{O}{iii}]  \\
 VVDS      & 140031  &   2.2141   & 0   &  2 &  3.5   &  H$\alpha$+[\ion{O}{iii}]  \\
 VVDS      & 142226  &   1.5165   & 0 .0001 & 2 &   3.5     &  H$\alpha$  \\
 VVDS      & 144923  &   1.5600     & 0.0001   &  2 &   3.5    &  H$\alpha$  \\
 VVDS      & 160224  &   1.6061   &  0  & 3 & 3.667 &   H$\alpha$  \\
 VVDS      & 162982  &   2.2442   & 0.0007    & 2 &   4.0      &  H$\alpha$+[\ion{O}{iii}]  \\
 VVDS      & 163515  &   1.5716   &  0.0001 & 3 & 3.5     &   H$\alpha$  \\
 VVDS      & 165659  &   2.1396   & 0.0001   & 2 &   3.5      &  H$\alpha$+[\ion{O}{iii}]  \\
 VVDS      & 165940  &   2.2649   & 0.0002  & 2 &   3.5     &  H$\alpha$  \\
 VVDS      & 167263  &   2.2599     & 0.0002  & 2 &   3.5      &  H$\alpha$  \\
 VVDS      & 167264  &   2.2591   & 0.001    & 2 &   3     &   [\ion{O}{iii}]  \\
 VVDS      & 168107  &   1.5599     & 0.0003  & 2 &   3      &  H$\alpha$  \\
 VVDS      & 168869  &   2.4433   & 0.0004  & 2 &   3.5    &  H$\alpha$+[\ion{O}{iii}]  \\
 VVDS      & 171656  &   2.2376   & 0.0001   & 2 &    3     &  H$\alpha$+[\ion{O}{iii}]  \\
 VVDS      & 179905  &   2.0605   & 0.0003  & 2 &    4    &  H$\alpha$  \\
 VVDS      & 179990  &   2.0356   & 0          & 2 &   3     &  H$\alpha$   \\
 VVDS      & 383107  &   2.1861   & 0           & 2 &   3      &  H$\alpha$+[\ion{O}{iii}]  \\
 VVDS      & 515718  &   2.3259   & 0.0003  & 2 &   3       &  H$\alpha$+[\ion{O}{iii}]  \\
 \hline
 \end{tabular}
\end{table*}

\section{Results}
\label{Sec:results}

\begin{figure*}[h!]
    \centering
    \includegraphics[scale=0.4]{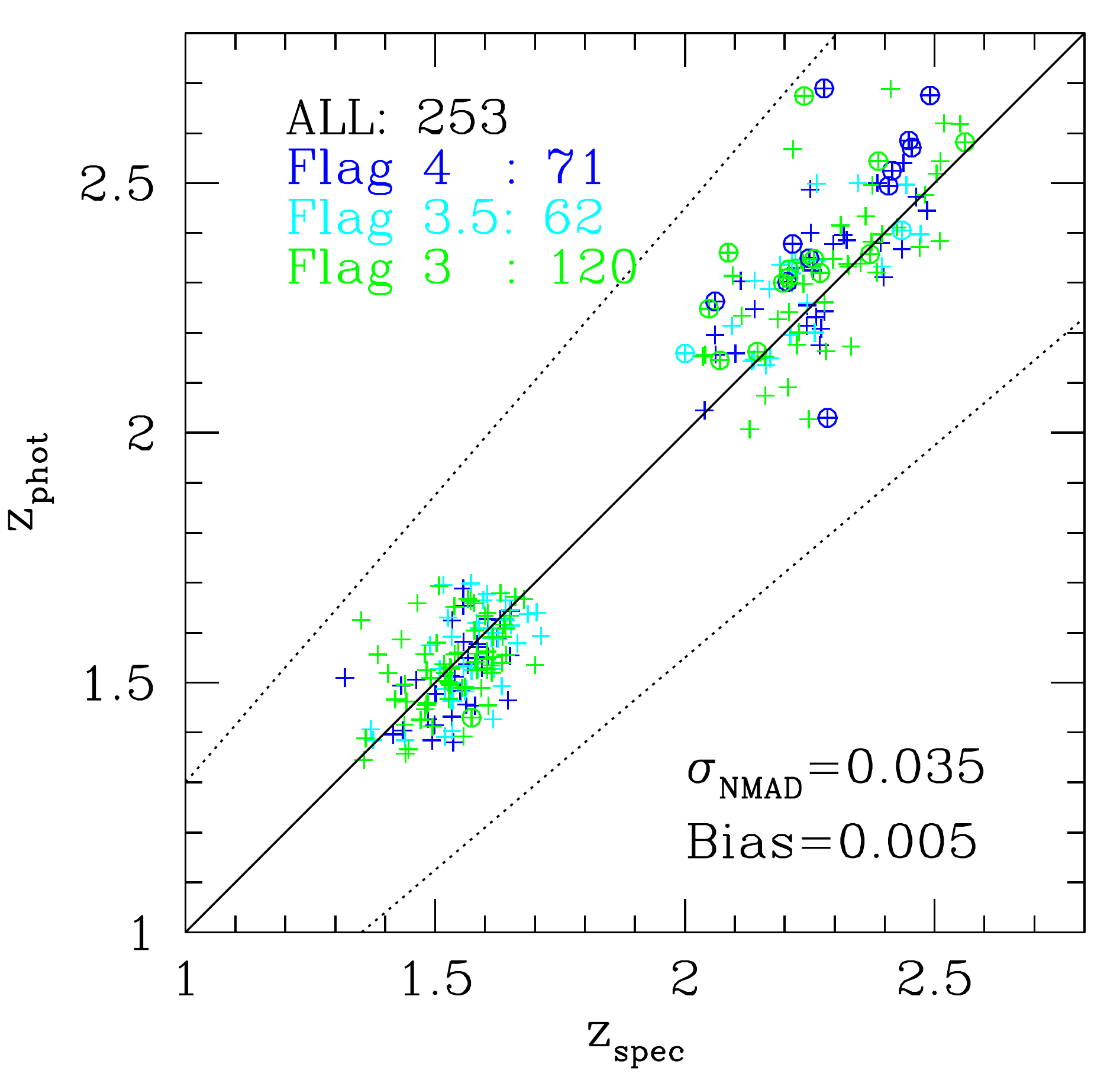}
     \includegraphics[scale=0.4]{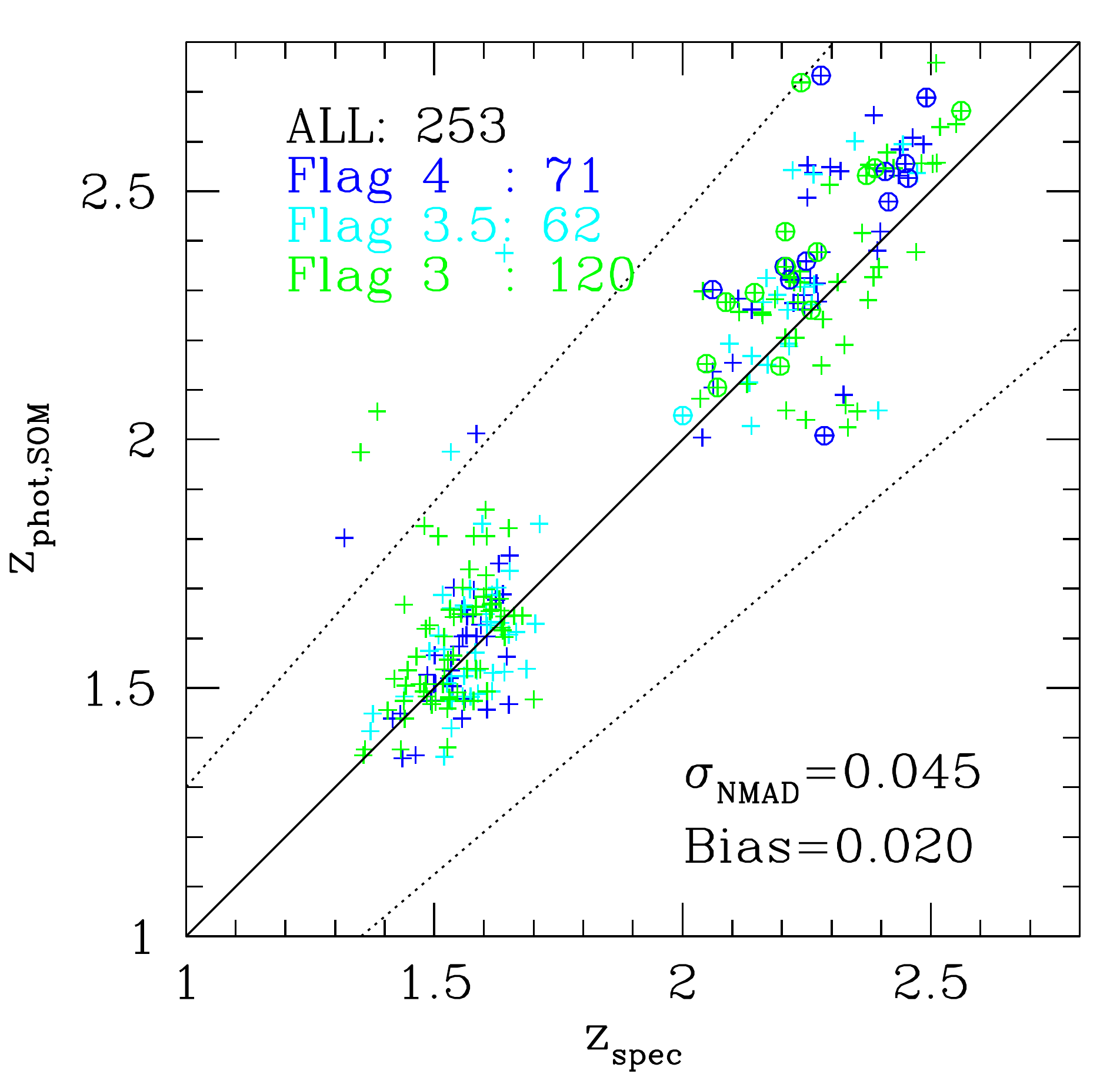}
    \caption{Comparison between photometric and spectroscopic redshifts. {\it Left}: Comparison between $z_{\rm phot}$ and $z_{\rm
        spec}$ for high-quality $({\rm Flag} \geq 3)$ redshift galaxies
      observed during the LBT campaign. Crosses show spectroscopic
      redshifts determined from the H$\alpha$ line, circles with
      crosses from the [\ion{O}{iii}] lines. The dotted lines define the
      region outside which the $z_{\rm phot}$ is considered a
      catastrophic failure,  defined by a redshift error
      $|z_{\rm phot}-z_{\rm spec}|/(1+z_{\rm spec}) \geq 0.15$.
      {\it Right}: Same as the {\it left} panel, but comparing $z_{\rm phot,SOM}$
      and $z_{\rm spec}$. The values of $\sigma_{NMAD}=1.48 {\rm Median}(|z-z_{\rm spec}|/(1+z_{\rm spec}))$ and ${\rm Bias}={\rm Mean}((z-z_{\rm spec})/(1+z_{\rm spec})))$ are also given, with $z=z_{\rm phot}$ in the left plot and  $z=z_{\rm phot,SOM}$ in the right.}
    \label{fig:zspec_vs_zphot}
\end{figure*}

We measured good $({\rm Flag} \geq 3)$ spectroscopic redshifts for 253
objects (two of which already known):   71 with ${\rm Flag} =4$, 62
with ${\rm Flag} =3.5$, and 120 with ${\rm Flag} =3$.  Figure
\ref{fig:zspec_vs_zphot} compares these to the photometric (left) and
the SOM-based (right) redshifts. The photometric redshifts are from
\citet{Ilbert2006} and \citet{Laigle2016}; the SOM-based redshifts are
the averages of the photometric redshifts of the galaxies belonging to
the SOM cell. There are no catastrophic failures with $|z_{\rm
  phot}-z_{\rm spec}|/(1+z_{\rm spec}) \geq 0.15$ and ${\rm Flag} \geq
3$.  Seven objects have $|z_{\rm phot,SOM}-z_{\rm spec}|/(1+z_{\rm
  spec}) \geq 0.15$; we examine the cells they belong to below. The
values of $\sigma_{NMAD}=1.48 {\rm Median}(|z-z_{\rm spec}|/(1+z_{\rm
  spec}))$ and ${\rm Bias}={\rm Mean}((z-z_{\rm spec})/(1+z_{\rm
  spec})))$ are comparable to what achieved in previous C3R2 releases;
for example G2020 quote $\sigma_{NMAD}=0.03$ and ${\rm Bias}=-0.003$
when considering $z=z_{\rm phot}$ and $\sigma_{NMAD}=0.044$ and ${\rm
  Bias}=0.027$ when considering $z=z_{\rm phot,SOM}$.

\begin{figure*}[h!]
    \centering
    \includegraphics[scale=0.4]{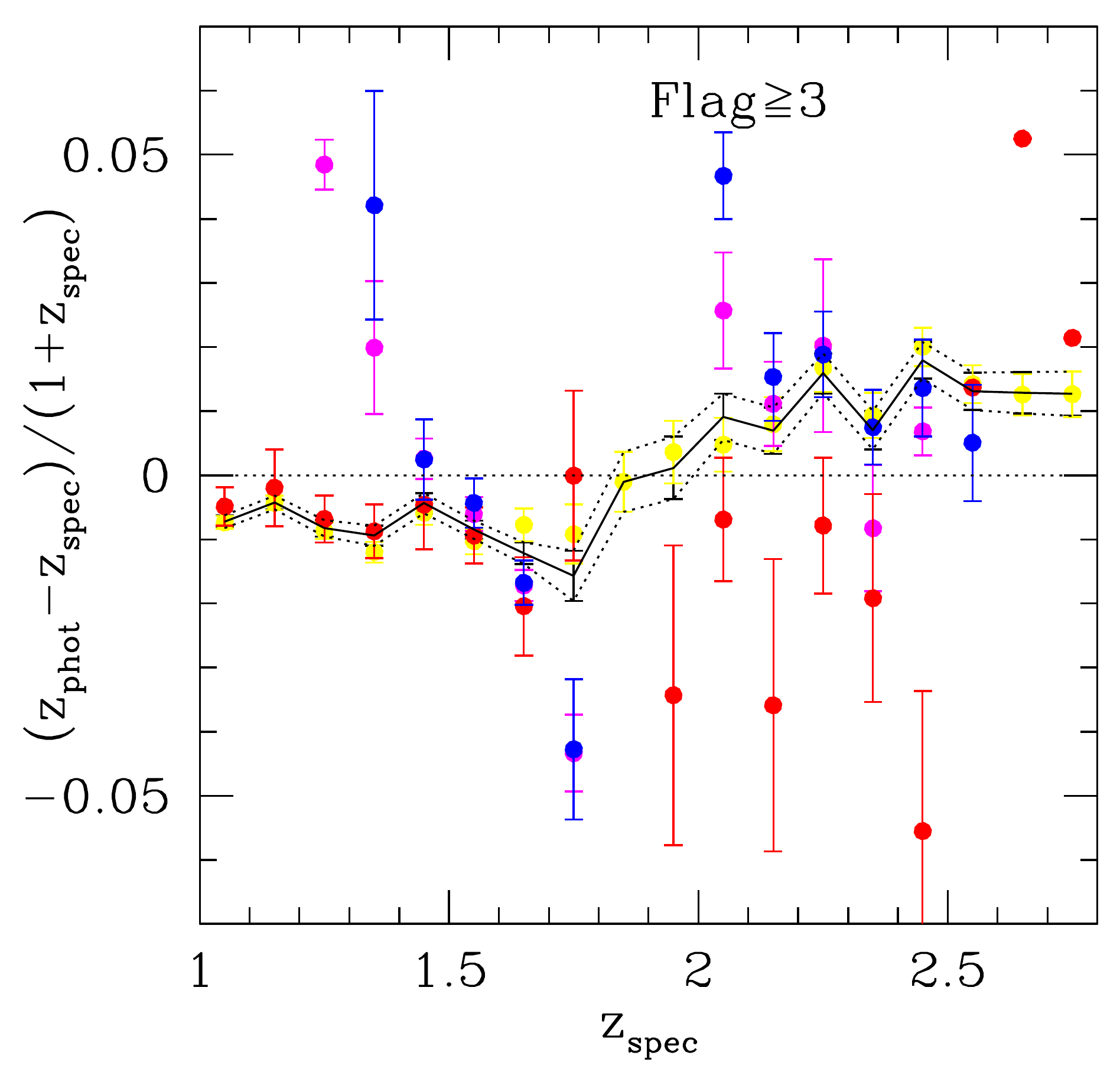}
     \includegraphics[scale=0.4]{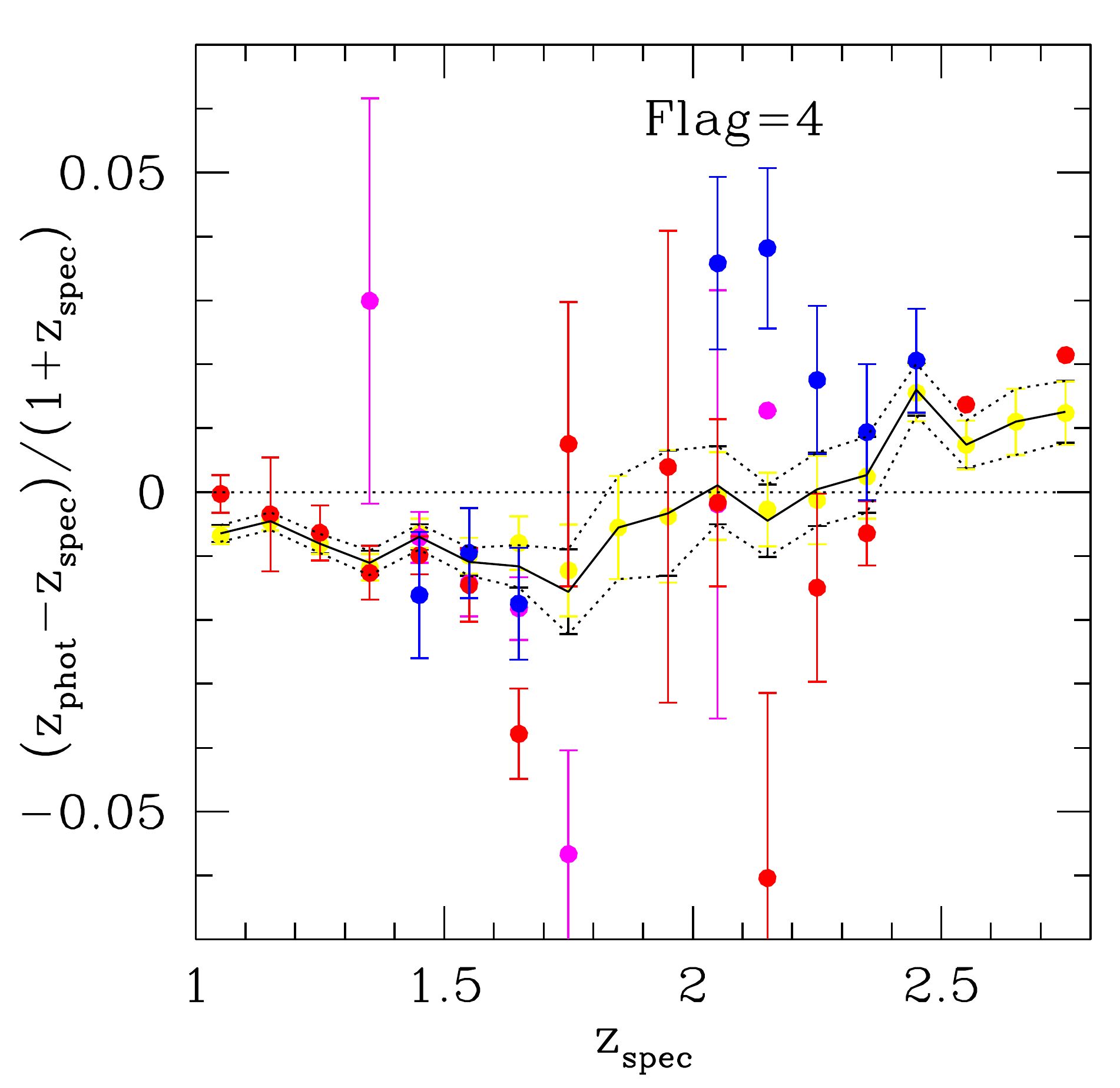}
     \caption{Mean differences between photometric and spectroscopic redshifts. {\it Left}:  Mean differences of
       $(z_{\rm phot}-z_{\rm  spec})/(1+z_{\rm spec}) $ of non-catastrophic failures
       in bins of $z_{\rm spec}$ as a
    function of    $z_{\rm spec}$ for ${\rm Flag} \ge3$ redshifts  released
    by the C3R2 project: yellow from \citet{Masters2015,Masters2017,Masters2019};
    magenta from G2020; red from
    \citet{Stanford2021}; blue, this paper. The black line shows the
    average in bins of the whole sample, the dotted lines the error range. 
    {\it Right}: Same, but  for ${\rm Flag}=4$ redshifts.}
    \label{fig:Allzspeczphot}
\end{figure*}

Figure \ref{fig:zspec_vs_zphot} shows that the agreement between
photometric and spectroscopic redshifts for objects with $z_{\rm
  spec}<2$ is excellent, with a mean difference of $-0.006$ and ${\rm
  RMS} =0.03$. At higher redshift, however, photometric redshifts
appear systematically larger, with $\langle (z_{\rm phot}-z_{\rm
  spec})/(1+z_{\rm spec})\rangle=0.02$ and similar RMS. We investigate
this issue by examining the whole C3R2  dataset published to date
(see Fig.  \ref{fig:Allzspeczphot}). When averaged in redshift bins of width
0.1, the mean difference $\langle (z_{\rm phot}-z_{\rm
  spec})/(1+z_{\rm spec})\rangle$ appears slightly negative ($\approx
-0.01\pm 0.0015$) up to redshifts 1.7, and slightly positive ($\approx
+0.01\pm 0.003$) at higher redshifts, consistently in all datasets and
for both ${\rm Flag} \ge3$ or ${\rm Flag} =4$ redshifts. Moreover, in
each bin up to redshift 1.7 the mean difference, or bias, is well
determined with a signal-to-noise ratio S/N between 3 and 8. This S/N
is lower (between 2 and 6) at redshifts higher than 2; the bias is
unconstrained (with S/N less than 1) in between. The S/N improvement in
each bin achieved by the new redshifts released here is $\le 0.5$. 

This shows that the C3R2
project is able to detect and correct the residual small biases
present in the most accurate available photometric redshift samples
that can be used as reference in the {\it Euclid} mission.  Therefore
the tight {\it Euclid} requirement that the mean redshift in each
tomographic bin must be constrained at the level of $0.002(1+z)$ is
achieved up to redshift 1.7 thanks to the calibration provided by the
C3R2 dataset. It is almost achieved for redshifts higher than 2, but
remains problematic in between, where the error on the possible
residual bias is $\approx 0.005$.

How the spectroscopic calibration of single SOM cells will be
implemented when building the lensing tomographic bins for {\it Euclid} is
still under investigation and goes beyond the scope of this paper.

\begin{figure*}
    \centering
    \includegraphics[scale=0.7,clip]{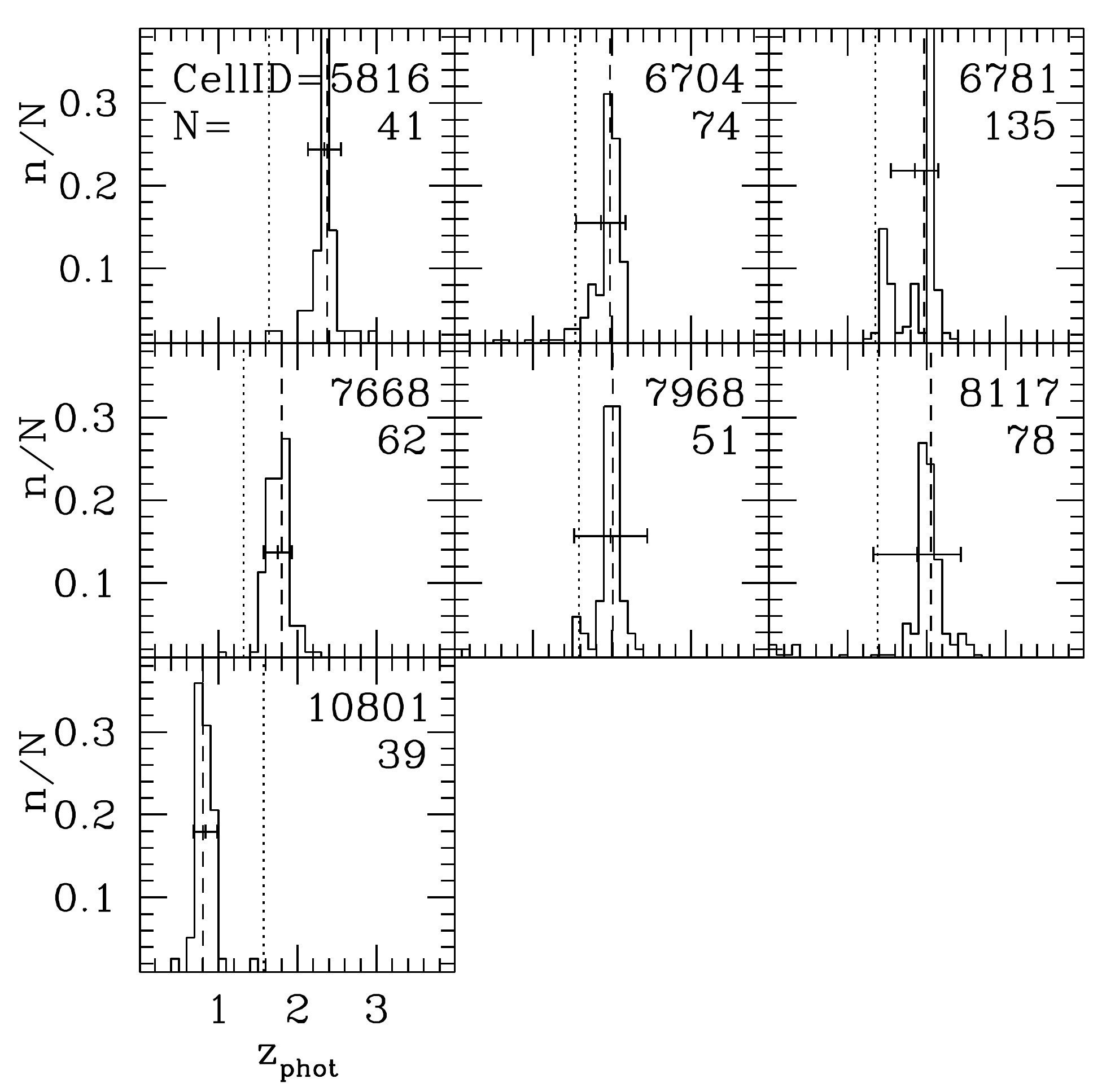}    
    \caption{Histogram of $z_{\rm phot}$ of galaxies populating each
      cell with $|z_{\rm phot,SOM}-z_{\rm spec}|/(1+z_{\rm
        spec})>0.15$ (right panel of
      Fig. \ref{fig:zspec_vs_zphot}). The distribution is normalized
      by dividing the number of galaxies in each $z_{\rm phot}$ bin by
      the total number of $z_{\rm phot}$ populating the considered
      cell; the number is indicated by the letter N in the top left
panel, and written in the same position in the
      others. Similarly, the cell number (CellID) is also given in each panel. The $z_{\rm phot,SOM}$ is represented by the dashed line, whereas dotted lines indicate $z_{\rm spec}$ measured during our program. The horizontal bar centered on the mean $z_{\rm phot}$ is the RMS of the histogram.}
    \label{fig:cell_occupation_zphot}
\end{figure*}

Figure \ref{fig:zspec_vs_zphot}, right shows that six galaxies have
$|z_{\rm phot,SOM}-z_{\rm spec}|/(1+z_{\rm spec})>0.15$. A seventh
galaxy at $z_{\rm phot,SOM}=0.8$ (and therefore not visible in the
figure) has a similarly discrepant redshift. As noted in G2020, they
belong to cells with a large spread in photometric redshifts, probing
the second peak of the distribution or its tail (see Fig.
\ref{fig:cell_occupation_zphot}).

\begin{figure}[h!]
    \centering
    \includegraphics[scale=0.45]{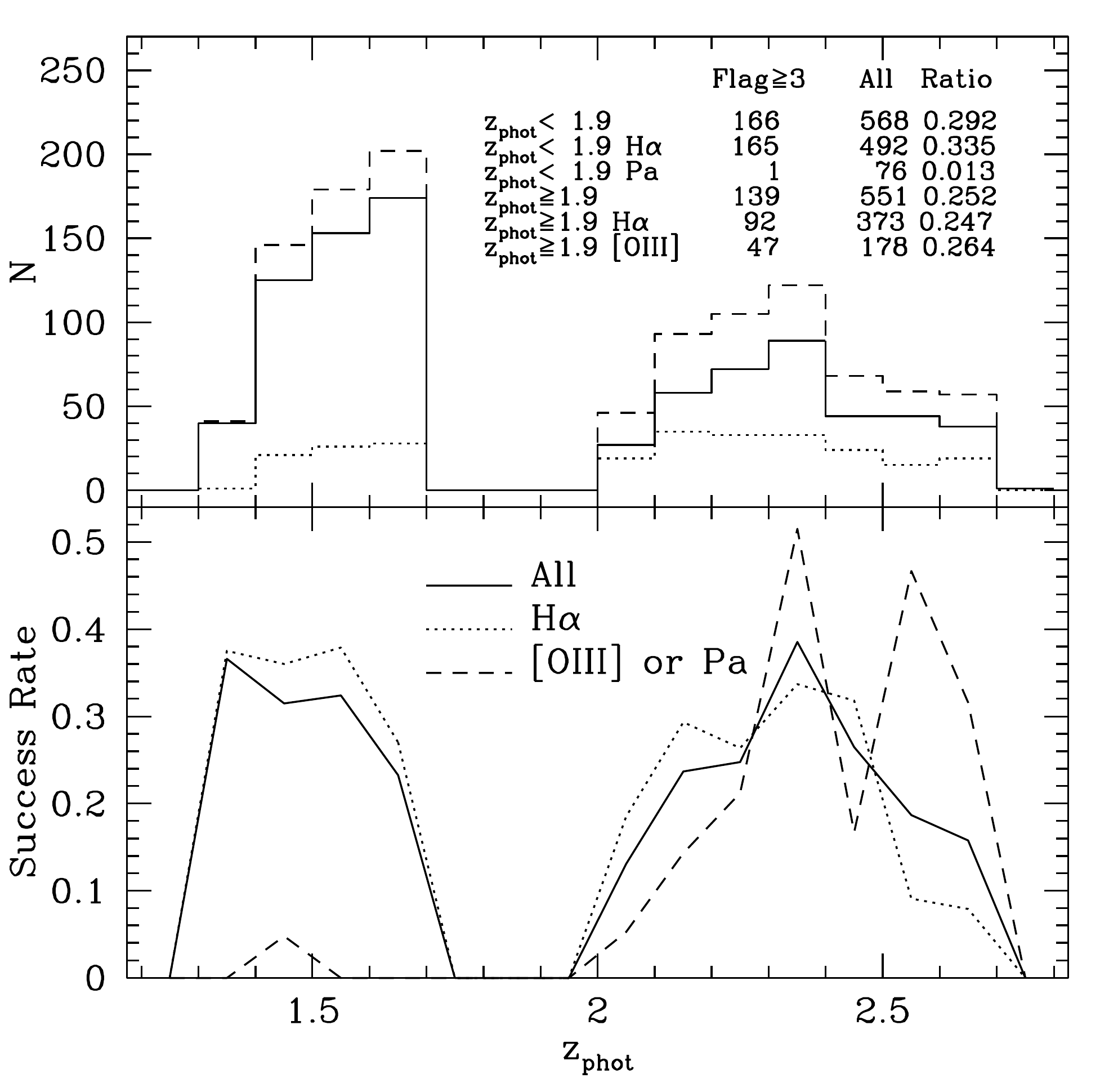}
    \caption{Spectroscopic success rate as a function of redshift.{\it Top}: Histogram of $z_{\rm phot}$ of targeted
      galaxies (dashed lines), targeting H$\alpha$   (solid lines), and the [\ion{O}{iii}] or Pa lines (dotted lines).  {\it
        Bottom:} Success rate as a function of $z_{\rm
        phot}$ considering all lines (H$\alpha$, [\ion{O}{iii}], Pa, full
      line), based on H$\alpha$ only (dotted line), or on [\ion{O}{iii}] or
      Pa (dashed line).}
    \label{fig:histoz}
\end{figure}

\begin{figure}[h!]
    \centering
   \includegraphics[scale=0.45]{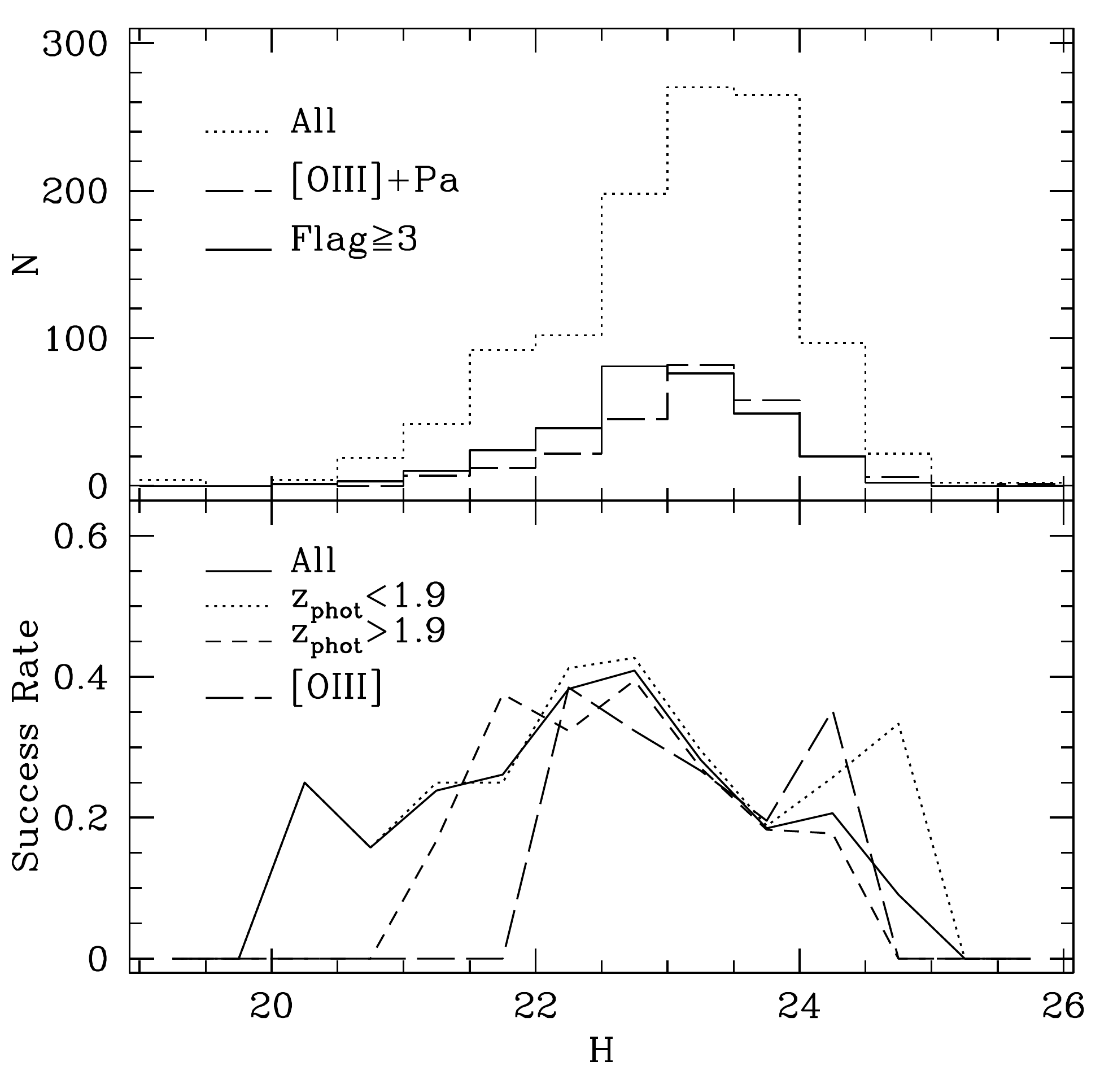}
   \caption{Spectroscopic success rate as a function of
       magnitude.
        {\it
        Top}: Histogram of the observed $H$ total magnitudes for
      all observed targets (dotted line), for secondary targets
      (long-dashed line), and for those with high-quality
      spectroscopic redshifts (full line).  {\it
        Bottom}: Success rate as a function of $H$ total
      magnitudes over the whole probed redshift range (full line), and
    split according to band (dotted: $H$ band or $1.3<z_{\rm phot}<1.7$;
    dashed: $K$ band, or  $2<z_{\rm phot}<2.7$. The long-dashed line
    shows the success rate targeting the [\ion{O}{iii}] line.}
    \label{fig:histoHmag}
\end{figure}

Figure \ref{fig:histoz} examines the success
rate as a function of redshift. 
As already discussed above, the success rate is lower in
the $K$ band and peaks when the redshifted H$\alpha$ line is around the
mid-redshift of either band. This is expected since the wavelength
coverage depends upon the position of the galaxies on the masks, and we
set the central wavelength of the spectrographs to the average of the
expected H$\alpha$ positions. Moreover, the focus of the spectrographs
deteriorates for objects that are not within the central 2.8\,arcmin
stripe of the field. Similarly, the [\ion{O}{iii}] lines at
$z_{\rm phot}\le 2.1$ fall outside of the covered wavelength range in
the $H$ band if the objects are not in extreme positions on the
masks. Overall, the success rate achieved for $z_{\rm phot}\ge 2.0$
objects targeting the H$\alpha$ in the $K$ band or the [\ion{O}{iii}] lines in
the $H$ band is similar.  In contrast, we managed to detect only once
(with ${\rm Flag}=3$)
the Pa lines from the 77 targeted  $z_{\rm phot}\le 1.7$ objects in the $K$ band. However,  these objects were selected as fillers when space was available in the
masks and no H$\alpha$ targets were left.
Figure \ref{fig:histoHmag} examines the success rate as a function of the
apparent total $H$-band magnitude of the galaxies.
This is not a strong function of the $H$ magnitude; it is
approximately 35\%, down to $H$ magnitudes of 23 and declining for fainter objects.

With the same motivation as in  G2020, we now examine the success rate
in populating SOM cells with
spectra. We targeted 518 cells (388 in the COSMOS field, 194 in the
VVDS field, with overlap), obtaining reliable spectra $({\rm Flag} \ge3)$
for 202 (127 in COSMOS, 94 in VVDS with overlap), translating to a
success rate of 39\% (33\% in COSMOS, 48\% in VVDS). Of the 202 cells
probed successfully here, 49 were empty before these observations;
for the remaining 153 cells with at least one good spectroscopic redshift from
previous C3R2 campaigns, on average our observations increased the
number of spectroscopic redshifts per cell by  72\%.

\begin{figure}[h!]
    \centering
   \includegraphics[scale=0.45]{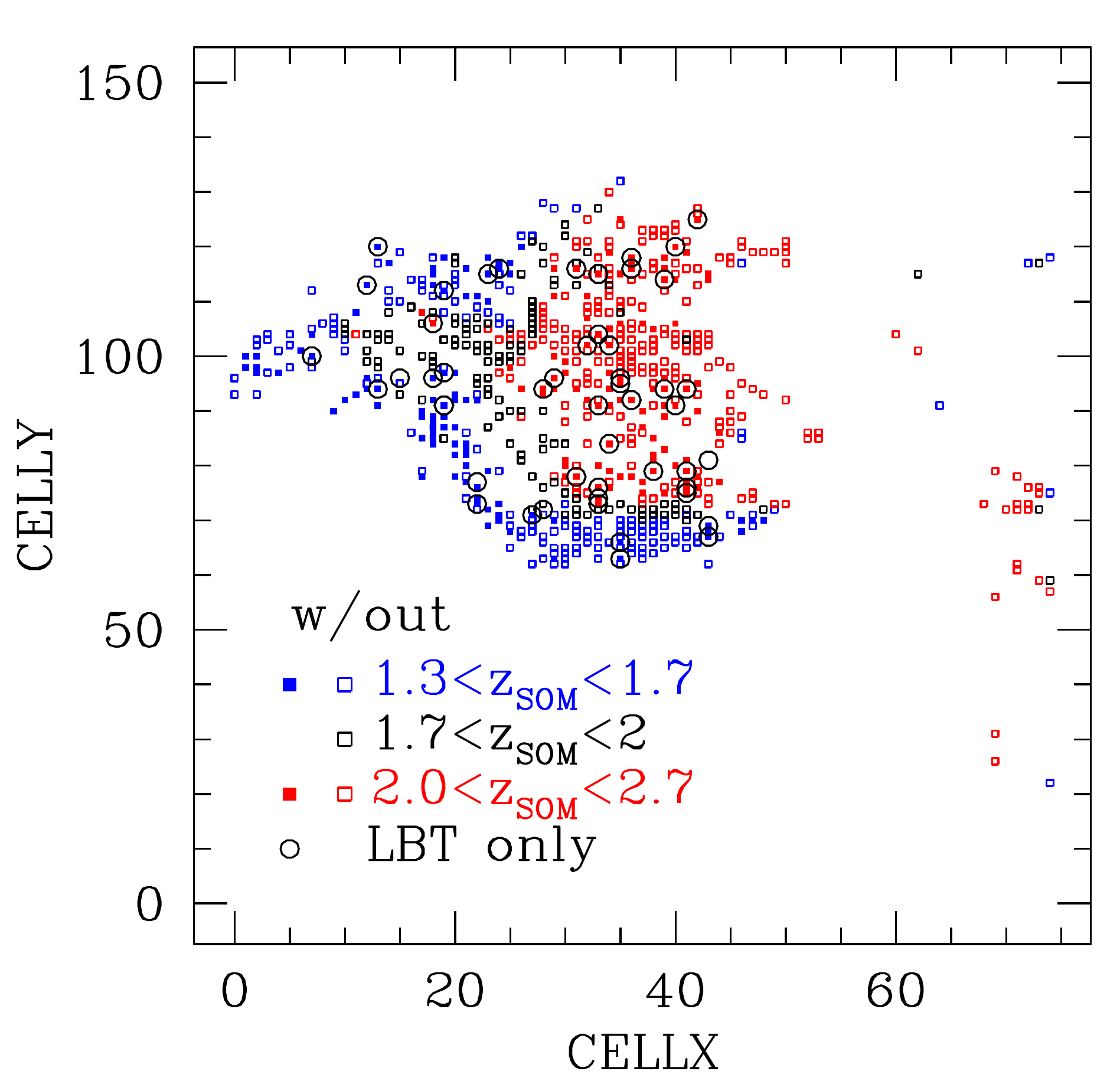}
    \caption{Cells probed by the current spectroscopic release (filled squares) and  cells still lacking spectroscopy (empty squares). Blue cells have $1.3<z_{\rm phot,SOM}<1.7$, black cells have $1.7<z_{\rm phot,SOM}<2.0$, and red cells have $2.0<z_{\rm phot,SOM}<2.7$. Open circles are the 49 cells without spectroscopy before this release. The circles without a centered square have $z_{\rm phot,SOM}$ outside the probed range.}
    \label{fig:SOM}
\end{figure}

Figure \ref{fig:SOM} shows the distribution of the SOM cells probed by
the current spectroscopic release, together with the cells still
empty. Summing up the contributions of all C3R2 releases, we find the
following. In the 839 cells with $1.3<z_{\rm phot,SOM}<1.7$, there are
52491 galaxies (with I-band magnitudes brighter than 24.5 in the COSMOS
and VVDS fields).  In 683 of these cells (81\%), where 45370 galaxies
are found (86\% of the total), we   collected at least one good
spectroscopic redshift. In the 895 cells with $2<z_{\rm
  phot,SOM}<2.7$, there are 40013 galaxies. In 683 of these cells
(74\%), where 30981 galaxies are found (77\% of the total), we  
collected at least one good spectroscopic redshift.

\section{Discussion and conclusions}
\label{Sec:conclusions}

We present the results of the LBT campaign to calibrate photometric
redshifts in the redshift range 1.3 to 2.7 using the LUCI near-infrared
spectrographs in the framework of the C3R2 project. We observed 88
masks, 58 in the COSMOS field and 30 in the VVDS field, with an
average of 12
objects per mask, aiming to detect  the H$\alpha$ line for primary
targets, and the  [\ion{O}{iii}] or Pa lines for secondary objects. We
extracted 1119 spectra, 1100 of which were identified as C3R2
galaxies. From 292 of these galaxies  we were able to measure reliable
spectroscopic redshifts. We assessed their precision from repeated
measurements to be in the range $\Delta z\sim 0.0002$ to 0.0004.
After averaging repeat
measurements, we ended up with reliable spectroscopic redshifts for 253
galaxies, two of which with already-known values.

Comparison with the C3R2 photometric redshifts shows that none of
these galaxies are catastrophic outliers. The values of $\sigma_{\rm NMAD}$ and
the bias are comparable to those reported in the previous C3R2
releases. Analyzing the whole C3R2 spectroscopic database published to date, we detect a small systematic shift $\langle (z_{\rm
  phot}-z_{\rm spec})/(1+z_{\rm spec})\rangle$ on the order of
$-0.01\pm 0.0015$ in the redshift range $1.3<z_{\rm phot}<1.8$ and of
$+0.01\pm 0.003$ in the range $2<z_{\rm phot}<2.7$. This is relevant
for the calibration of the redshifts of SOM cells in these redshift
ranges, essentially matching the {\it Euclid} requirement of knowing
the mean redshift of each tomografic bin to $0.002(1+z)$.  Our
redshift determinations populate 49 SOM cells with no prior
spectroscopic measurements, approximately doubling the occupation
numbers of an additional 153 cells. Seven SOM cells have $|z_{\rm
  phot,SOM}-z_{\rm spec}|/(1+z_{\rm spec}) \geq 0.15$; they have a
large spread in photometric redshifts, and our objects probe the
second peak of the distribution or its tail.

In the redshift range 1.3 to 1.7 there are still 156 cells (19\%)
without at least one good spectroscopic redshift; in these cells that
lack a spectroscopic calibration we find only 14\% of the galaxy
population. In the redshift range 2 to 2.7 there are  232 cells (26\%)    without at least
one good spectroscopic redshift; in these cells we find
23\% of the galaxy population.  In between, in the
redshift range 1.7 and 2 where telluric absorption makes the
direct observation of H$\alpha$ impossible, the number of cells without at
least one good spectroscopic redshift is 126 (28\%); in these cells we find 
26\% of the galaxy population with redshifts between 1.7 and 2. The redshift bias
$\langle (z_{\rm phot}-z_{\rm spec})/(1+z_{\rm spec})\rangle$ is not well constrained, with an error (0.005) larger than the   {\it Euclid} requirement.

Two questions arise.  First, is it worth attempting to calibrate the missing
cells in the $H$ and $K$ bands with LUCI at the LBT? Probably not, since
 their density is too low for the field of view available. Moreover,
G2020 found that most of them have low star formation
rates, making the detection of emission lines very difficult, and the
success rate (around 30\%) of the present campaign is not particularly
encouraging. Second, is it worth attempting the calibration of the cells missing in the
redshift range 1.7 to 2 with ground-based facilities? Detecting the H$\gamma$, H$\beta$, and [\ion{O}{iii}] lines in the
$J$ band up to redshift 2, 1.9, and 1.8, respectively, could be possible with an
increasing probability of success. Whether the LBT with the LUCI
spectrographs should be used for such a campaign depends on the
density on the sky of the objects needing spectra. However, the
difficult remaining sources at these redshifts are primarily passive
galaxies, and so spectroscopic searches for Balmer and [\ion{O}{iii}]
emission lines are unlikely to be efficient or successful. A costly
solution could be ground-based (optical) absorption-line
spectroscopy. A release of optical spectra gathered at the VLT and the
GRANTECAN telescopes is in preparation. In the end the (possibly
partial) natural solution will come from {\it Euclid} itself: its
spectroscopic program will deliver near-infrared spectra in the
relevant redshift range free of telluric absorption. The survey is
designed to detect emission lines down to $2\times10^{-16} {\rm erg s}
^{-1} {\rm cm}^{-2}$ at $3.5\sigma$ in the wide configuration, and
reach up to 2 mag fainter levels in the deep fields
\citep{Scaramella2021}, or $3$ to $6\times 10^{-17} {\rm erg s}^{-1}
      {\rm cm}^{-2}$. Using the H$\alpha$ fluxes measured by G2020, we
      estimate that 60 to 90\% of the sources will be detected within
      these limits in the deep fields. Therefore, the mission itself
      will provide enough spectra to finish the photometric redshift
      calibration in the near-infrared range to the required
      precision. In particular, a question that needs to be clarified is
      whether in a given SOM cell the galaxies for which we are able or
      unable to measure a spectroscopic redshift might have different
      redshift distributions.

\begin{acknowledgements}
\AckEC

RS, MF, VG, RB acknowledge support by the Deutsches Zentrum f\"ur 
Luft- und Raumfahrt (DLR) grant 50 QE 1101. The work of DS was carried
out at the Jet Propulsion Laboratory, California Institute of
Technology, under a contract with NASA. We thank the staff of the LBT
observatory for their support during the mask preparation and the
execution of the observations.

\end{acknowledgements}

\bibliographystyle{aa}
\bibliography{bibliography.bib}

\begin{appendix}
\section{Table A.1}
\begin{table*}
    \caption{List of the observed masks.}
    \label{Tab:observed_pointings}
\begin{tiny}
\begin{tabular}{lccccccccc}
\hline
Field & Mask ID & RA$_{\rm cen}$ & Dec$_{\rm cen}$ & Rot.Angle & nExp
      & UT  Date & JD  & $z_{\rm helcorr}$ & Success Rate \\
        &                & (hms)        & (dms)           &  (deg) & $(\times  200\,{\rm  s}$)
   &(yyyy.mm.dd)  &&$(\times  10^5)$&  $({\rm Flag} \geq 3/{\rm Prim}/{\rm Obs})$\\
  \hline
VVDS   & M1H     & 022614.372 & $-045554.320$ &  15 & 36 & 2019-10-03 & 2458759.83&3.6 & 6/12/14\\ 
VVDS   & M2H   & 022622.493 & $-045356.990$ &  20 & 36 & 2019-10-03 & 2458759.95&3.5 & 7/8/12\\ 
VVDS   & M3H & 022622.572 & $-045759.610$ & $-16$ & 36 & 2019-10-07 & 2458763.80&3.0 & 4/7/12\\ 
VVDS   & M4H     & 022606.238 & $-045557.250$ &   0 & 36 & 2019-10-07 & 2458764.93&2.7 & 4/10/14\\ 
VVDS   & M5H  & 022630.656 & $-044152.010$ &  75 & 36 & 2019-10-09 & 2458765.82&2.7 & 7/7/12\\ 
VVDS   & M9H     & 022550.162 & $-045358.310$ &  50 & 22 & 2020-01-19 & 2458867.75&$-9.7$ & 6/9/13\\
\hline
  \end{tabular}
\end{tiny}
\end{table*}
\setcounter{table}{0}
    \begin{table*}[h!]
    \caption{continued.}
\begin{tiny}
    \begin{tabular}{lccccccccc}
\hline
Field & Mask ID & RA$_{\rm cen}$ & Dec$_{\rm cen}$ & Rot.Angle & nExp & UT Date & JD&$z_{\rm helcorr}$ & Success Rate \\
        &                & (hms)        & (dms)           &  deg & $(\times 200\,{\rm s})$    &(yyy.mm.dd)  &&$(\times   10^5)$&$({\rm Flag}\geq 3/{\rm Prim}/{\rm Observed})$\\
  \hline
COSMOS & M23H  & 095912.153 & +015008.170 &   90 & 23 & 2021-03-21 &2459294.15&$-4.6$ & 3/11/15 \\ 
COSMOS & M23K  & 095912.153 & +015008.170 &   90 & 23  & 2021-03-21 &2459294.15 &$-4.6$ & 1/8/12  \\
COSMOS & LAE3H & 100025.092 & +021627.160 &    0 & 19  & 2021-03-22 & 2459295.19&$-4.7$ & 0/6/6 \\ 
COSMOS & LAE3K  & 100025.092 & +021627.160 &    0 & 19  & 2021-03-22 &2459295.19 &$-4.7$ & 0/4/5 \\
COSMOS & LAE4H  & 100041.352 & +021239.210 &  $-15$ & 36  & 2021-03-23 &2459296.18 &$-4.9$ & 0/3/6 \\ 
COSMOS & LAE4K  & 100041.352 & +021239.210 &  $-15$ & 36  & 2021-03-23 &2459296.18 &$-4.9$ & 3/6/10 \\
\hline
  \end{tabular}
\end{tiny}
\end{table*}
\FloatBarrier

\section{Table B.1}
\begin{table*}
\begin{center}
  \caption{Sample of galaxies with good spectroscopic redshifts.
    The explanation of the different columns is given in
    Sect. \ref{sec:redshift_assignment}. }

  \label{Tab:redshifts}
\begin{tiny}

\begin{tabular}{lrrrrrr}
\hline
  \multicolumn{1}{c}{OBJ\_ID} &
  \multicolumn{1}{c}{RA} &
  \multicolumn{1}{c}{Dec} &
  \multicolumn{1}{c}{$z_{\rm spec}$} &
  \multicolumn{1}{c}{Flag} &
  \multicolumn{1}{c}{Line} &
  \multicolumn{1}{c}{1D-Spectrum} \\
  \hline
       13275            & 150.3162252           &   1.7379527           &       2.4078         &         4.0           &      [\ion{O}{iii}]           & COSMOS\_LBT\_JAN21\_1d\_13275\_M13Hsky\_S21          \\
       20825            & 150.0527463           &   1.7531398           &       1.6520         &         4.0           &          H$\alpha$           & COSMOS\_LBT\_FEB21\_1d\_20825\_M20Hsky\_S08          \\
       21075            & 150.2729757           &   1.7538638           &       1.5172         &         3.0           &          H$\alpha$           & COSMOS\_LBT\_JAN21\_1d\_21075\_M13Hsky\_S13          \\
       21964            & 150.0395616           &   1.7555454           &       2.5610         &         3.0           &      [\ion{O}{iii}]           & COSMOS\_LBT\_FEB21\_1d\_21964\_M20Hsky\_S19          \\
       26585            &  150.079045            &   1.7650675           &       2.3980          &         4.0           &          H$\alpha$           & COSMOS\_LBT\_FEB21\_1d\_26585\_M20Ksky\_S16          \\
       28028            & 150.3115341           &   1.7680673           &       1.5573         &         3.5           &          H$\alpha$           & COSMOS\_LBT\_JAN21\_1d\_28028\_M13Hsky\_S17          \\
       31318            & 150.0632315           &   1.7735528           &       2.2090         &         3.0           &          H$\alpha$           & COSMOS\_LBT\_FEB21\_1d\_31318\_M20Ksky\_S11          \\
\hline
\end{tabular}
 \end{tiny}
  \end{center}
\end{table*}

\setcounter{table}{0}
  \begin{table*}
\begin{center}
  \caption{continued. }

\begin{tiny}

\begin{tabular}{lrrrrrr}
\hline
  \multicolumn{1}{c}{OBJ\_ID} &
  \multicolumn{1}{c}{RA} &
  \multicolumn{1}{c}{Dec} &
  \multicolumn{1}{c}{$z_{\rm spec}$}  &
  \multicolumn{1}{c}{Flag} &
  \multicolumn{1}{c}{Line} &
  \multicolumn{1}{c}{1D-Spectum} \\
  \hline
      311645           & 149.7982448           &   2.3368401           &        1.6048      &         3.0           &          H$\alpha$           & COSMOS\_LBT\_APR21\_1d\_311645\_M27Hsky\_S15          \\
      315750           & 150.3914582           &   2.3435241           &        1.4790      &         3.0           &          H$\alpha$           & COSMOS\_LBT\_FEB21\_1d\_315750\_M14Hsky\_S09          \\
      320989           & 149.8092325           &   2.3544984           &        2.3707      &         3.0           &      [\ion{O}{iii}]           & COSMOS\_LBT\_APR21\_1d\_320989\_M27Hsky\_S21          \\
      321846           & 150.3609407           &   2.3563503           &        1.5855      &         3.5           &          H$\alpha$           & COSMOS\_LBT\_FEB21\_1d\_321846\_M18Hsky\_S21          \\
      323850           & 149.9217085           &   2.3597902           &        2.1015      &         4.0           &          H$\alpha$           & COSMOS\_LBT\_FEB21\_1d\_323850\_M22Ksky\_S09          \\
      326545           & 150.3990555           &   2.3652568           &        1.4830      &         3.0           &          H$\alpha$           & COSMOS\_LBT\_FEB21\_1d\_326545\_M14Hsky\_S08          \\
      328061           & 150.3336896           &   2.3676283           &        1.6350      &         3.0           &          H$\alpha$           & COSMOS\_LBT\_FEB21\_1d\_328061\_M18Hsky\_S14          \\
\end{tabular}

 \end{tiny}
 
 \end{center}
\end{table*}
 
\setcounter{table}{0}
  \begin{table*}
\begin{center}
  \caption{continued. }

\begin{tiny}

\begin{tabular}{lrrrrrr}
\hline
  \multicolumn{1}{c}{OBJ\_ID} &
  \multicolumn{1}{c}{RA} &
  \multicolumn{1}{c}{Dec} &
  \multicolumn{1}{c}{$z_{\rm spec}$} &
  \multicolumn{1}{c}{Flag} &
  \multicolumn{1}{c}{Line} &
  \multicolumn{1}{c}{1D-Spectum} \\
  \hline
      494350           & 150.4846551               &   2.6939799             &    1.6412        &         3.0           &          H$\alpha$           & COSMOS\_LBT\_DEC20\_1d\_494350\_M6Hsky\_S14          \\
      549854           & 149.907150209           & 2.137621434           &      1.4802    &         3.0           &          H$\alpha$           & COSMOS\_LBT\_APR21\_1d\_549854\_M25Hsky\_S10          \\
      603116           & 150.197082912           &  2.21687046            &     2.4638     &         4.0           &          H$\alpha$           & COSMOS\_LBT\_MAR21\_1d\_603116\_LAE4Ksky\_S19          \\
      621780           & 150.471209414           &   2.2446394             &     2.4847      &         4.0           &          H$\alpha$           & COSMOS\_LBT\_APR21\_1d\_621780\_M24Ksky\_S08          \\
      646613           & 150.468731703           & 2.281641135           &      1.4908    &         3.0           &          H$\alpha$           & COSMOS\_LBT\_APR21\_1d\_646613\_M24Hsky\_S12          \\
      703468           & 149.78839164             &  2.36554692            &     2.3277      &         3.0           &          H$\alpha$           & COSMOS\_LBT\_APR21\_1d\_703468\_M27Ksky\_S09          \\
      720513           & 149.803548302           & 2.391162136           &      2.2512    &         4.0           &          H$\alpha$           & COSMOS\_LBT\_APR21\_1d\_720513\_M27Ksky\_S13          \\
\end{tabular}

 \end{tiny}
 
 \end{center}
\end{table*}
 
  \setcounter{table}{0}
  \begin{table*}
\begin{center}
  \caption{continued. }

\begin{tiny}

\begin{tabular}{lrrrrrr}
\hline
  \multicolumn{1}{c}{OBJ\_ID} &
  \multicolumn{1}{c}{RA} &
  \multicolumn{1}{c}{Dec} &
  \multicolumn{1}{c}{$z_{\rm spec}$} &
  \multicolumn{1}{c}{Flag} &
  \multicolumn{1}{c}{Line} &
  \multicolumn{1}{c}{1D-Spectum} \\
  \hline
      160224           &  36.6517582           &  $-4.6828961$           &        1.6061        &         3.0           &          H$\alpha$           & VVDS\_LBT\_OCT20\_1d\_160224\_M5Hsky\_S11          \\
      160224           &  36.6517582           &  $-4.6828961$           &        1.6061        &         4.0           &          H$\alpha$           & VVDS\_LBT\_OCT20\_1d\_160224\_M14Hsky\_S12          \\
      160224           &  36.6517582           &  $-4.6828961$           &        1.6061        &         4.0           &          H$\alpha$           & VVDS\_LBT\_OCT19\_1d\_160224\_M5Hsky\_S11          \\
      162982           &  36.6420709           &  $-4.6787762$           &        2.2448        &         4.0           &          H$\alpha$           & VVDS\_LBT\_OCT20\_1d\_162982\_M5Ksky\_S12          \\
      162982           &  36.6420709           &  $-4.6787762$           &        2.2435        &         4.0           &      [\ion{O}{iii}]           & VVDS\_LBT\_OCT19\_1d\_162982\_M5Hsky\_S20          \\
      163515           &  36.6101764           &  $-4.6765857$           &        1.5716        &         4.0           &          H$\alpha$           & VVDS\_LBT\_OCT20\_1d\_163515\_M5Hsky\_S16          \\
      163515           &  36.6101764           &  $-4.6765857$           &        1.5716        &         3.0           &          H$\alpha$           & VVDS\_LBT\_OCT20\_1d\_163515\_M14Hsky\_S15          \\
\end{tabular}

 \end{tiny}
 
 \end{center}
\end{table*}

\end{appendix}

\end{document}